\documentclass[12pt]{spieman}  
\usepackage{amsmath,amsfonts,amssymb}
\usepackage{graphicx}
\usepackage{setspace}
\usepackage{tocloft}
\usepackage{subcaption}
\usepackage{multirow}
\usepackage{diagbox}
\usepackage{lineno}

\title{Automatic Classification of Symmetry of Hemithoraces in Canine and Feline Radiographs }

\author[a, *]{Peyman Tahghighi}
\author[b]{Nicole Norena}
\author[a]{Eran Ukwatta}
\author[b]{Ryan B Appleby}
\author[c]{Amin Komeili}
\affil[a]{School of Engineering, University of Guelph, Guelph, Ontario}
\affil[b]{Department of Clinical Studies, Ontario Veterinary College, University of Guelph, Guelph, Ontario}
\affil[c]{Department of Biomedical Engineering, University of Calgary, Calgary, Alberta}

\cftpagenumbersoff{figure}
\cftpagenumbersoff{table} 
\begin{document} 
\maketitle

\begin{abstract}

\noindent
\textbf{Purpose}: Thoracic radiographs are commonly used to evaluate patients with confirmed or suspected thoracic pathology. Proper patient positioning is more challenging in canine and feline radiography than in humans due to less patient cooperation and body shape variation. Improper patient positioning during radiograph acquisition has the potential to lead to a misdiagnosis. Asymmetrical hemithoraces are one of the indications of obliquity for which we propose an automatic classification method. 

\noindent
\textbf{Approach}: We propose a hemithoraces segmentation method based on Convolutional Neural Networks (CNNs) and active contours. We utilized the U-Net model to segment the ribs and spine and then utilized active contours to find left and right hemithoraces. We then extracted features from the left and right hemithoraces to train an ensemble classifier which includes Support Vector Machine, Gradient Boosting and Multi-Layer Perceptron. Five-fold cross-validation was used, thorax segmentation was evaluated by Intersection over Union (IoU), and symmetry classification was evaluated using Precision, Recall, Area under Curve and F1 score.

\noindent
\textbf{Results}: Classification of symmetry for 900 radiographs reported an F1 score of $82.8\%$. To test the robustness of the proposed thorax segmentation method to underexposure and overexposure, we synthetically corrupted properly exposed radiographs and evaluated results using IoU. The results showed that the model’s IoU for underexposure and overexposure dropped by $2.1\%$, $1.2\%$ and, respectively.

\noindent
\textbf{Conclusions}: Our results indicate that the proposed thorax segmentation method is robust to poor exposure radiographs. The proposed thorax segmentation method can be applied to human radiography with minimal changes.

\end{abstract}


{\noindent \footnotesize\textbf{*}Peyman Tahghighi,  \linkable{ptahghig@uoguelph.ca} }

\begin{spacing}{2}   

\section{Introduction}

\label{sect:intro}  
Thoracic radiography is an important diagnostic  in companion animal medicine, routinely used for diagnosing cardiopulmonary disorders as well as the staging of neoplasms in dogs and cats. Radiographs must be of diagnostic quality to provide clinical utility to the veterinarian. The quality of the radiographs is heavily dependent on operator experience in acquisition and patient-related factors. Positioning errors are common, especially those where the patient is rotated \cite{2}.  In the authors’ experience, radiographs from general practices are inappropriately positioned approximately 30$\%$ of the time. Correcting positioning errors during image acquisition can improve diagnostic quality and is considered to be a key element in reducing errors in radiograph interpretation. While it is common practice to inspect  the image quality visually, the day-to-day challenges sometimes make this difficult to perform effectively. Therefore, a method for digital positioning error recognition would be valuable in clinical veterinary practice.

\begin{figure}[htb]
    \centering
    \begin{subfigure}[b]{\textwidth}
        \centering
        \includegraphics[width=0.3\linewidth, height=2in]{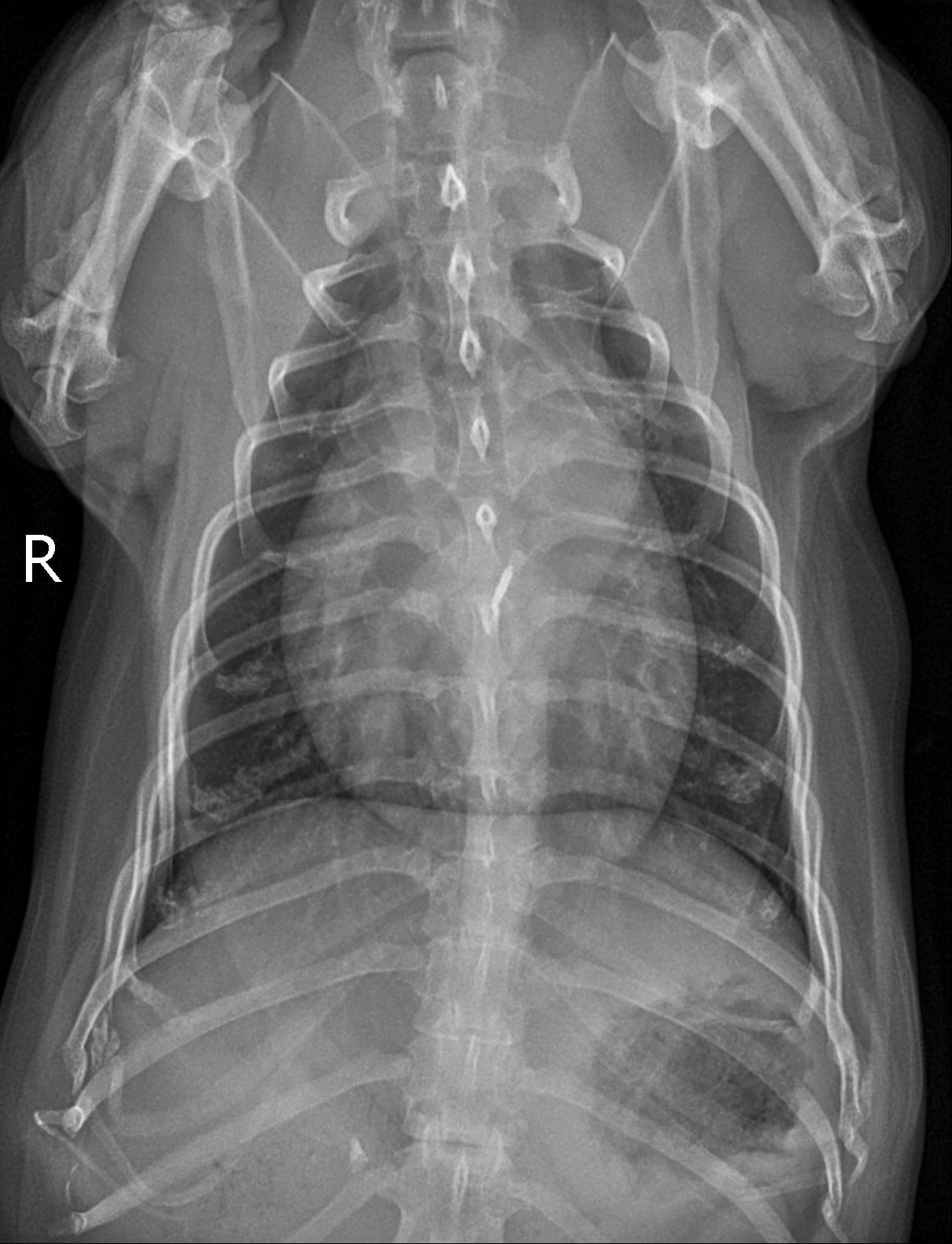}%
        \hspace{0.5in}
        \includegraphics[width=0.3\linewidth, height=2in]{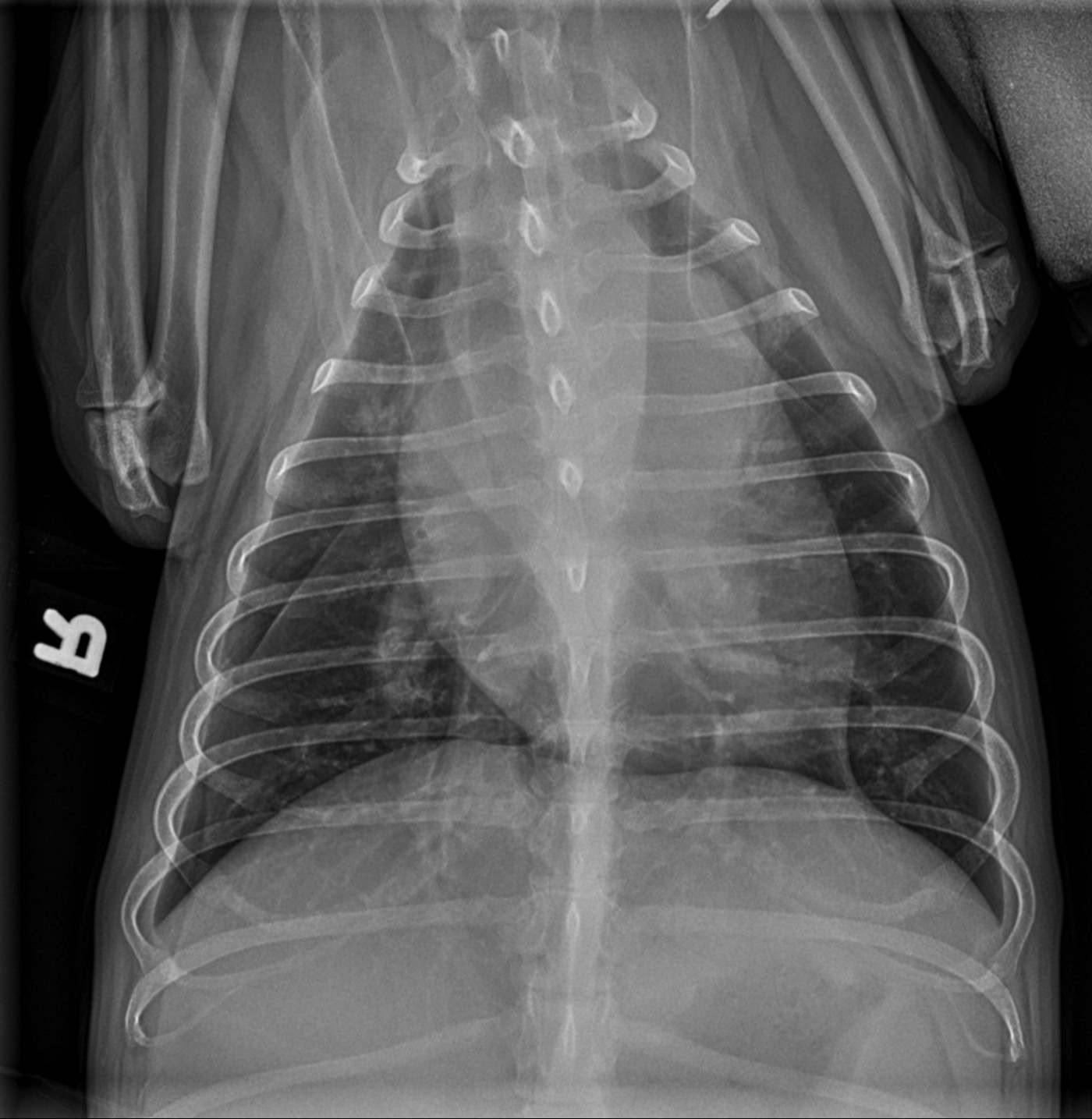}
        \caption{Symmetric}
        \label{fig1-1}
    \end{subfigure}
    \vskip\baselineskip
    \begin{subfigure}[b]{\textwidth}
        \centering
        \includegraphics[width=0.3\linewidth, height=2in]{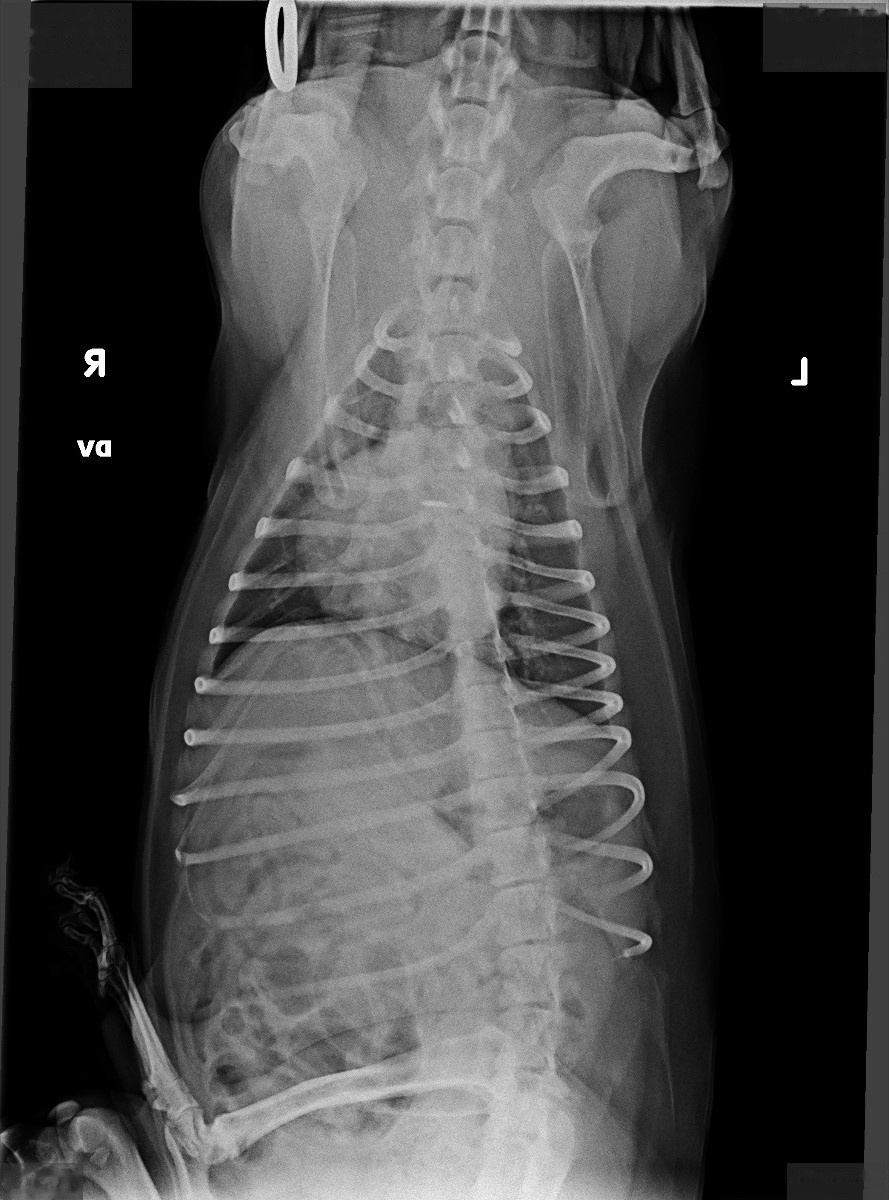}%
        \hspace{0.5in}
        \includegraphics[width=0.3\linewidth, height=2in]{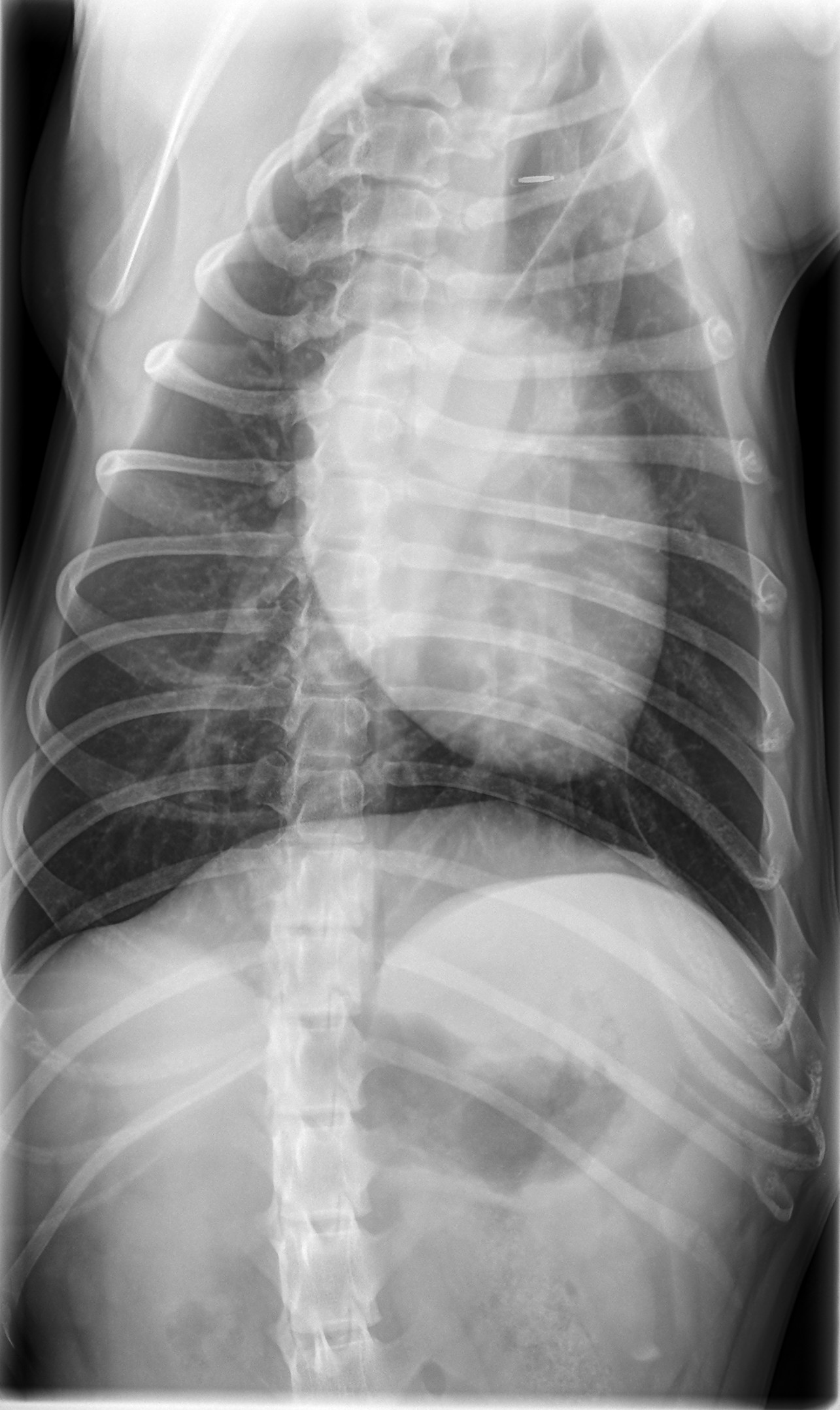}
        \caption{Asymmetric}
        \label{fig1-2}
    \end{subfigure}
    \caption{Examples of symmetric (a) and asymmetric hemithoraces (b) in canine and feline radiographs.}
    \label{fig1}
\end{figure}

The ventrodorsal (VD) and dorsoventral (DV) radiographs are  common projections of the thorax and are acquired in all thoracic radiograph studies. Patient positioning is aimed at collimating the animal so that the anatomy of the target organ is accurately captured in the radiograph. In a properly positioned VD or DV radiograph, the hemithoraces look nearly symmetrical \cite{1,2}, and severe deviation from this symmetry indicates that the patient has rotated during the positioning process. Fig. \ref{fig1} illustrates symmetric and asymmetric radiographs. In the asymmetric radiographs Fig. \ref{fig1-2}, the shape and size of the left and right hemithoraces highly differ, whereas, in the symmetric radiographs Fig. \ref{fig1-1}, these two regions are almost identical.

Deep learning methods have been extensively applied to canine radiographs in recent years for classification tasks. For instance, Yoon et al. \cite{3} used the Bag of Words combined with Convolutional Neural Networks (CNNs) to detect normal and abnormal canine thoracic radiographs, and used cardiomegaly as a metric of abnormality. Tonia et al. \cite{4} proposed an automatic sorting of canine radiographs by view and region. They heavily considered prediction time and memory limits in their work and proposed a novel lightweight CNN architecture based on ideas from SqueezeNet \cite{5} and Depthwise CNNs \cite{6}. Most recently, Banzato et al. \cite{7} utilized DenseNet-121 \cite{8}, and ResNet-50 \cite{9} to classify the lateral thoracic radiographs into 8 different findings such as cardiomegaly, alveolar pattern and bronchial pattern. In their work, they removed radiographs with positioning errors or poor image quality and used the whole radiograph as the input to the network without any prior segmentation. ResNet-50 achieved better classification results and for four classes such as cardiomegaly and pleural effusion, the Area Under Curve (AUC) of their model was similar to or higher than that reported in similar studies on humans. 

The problem of lung symmetry detection has already been studied in human chest radiography \cite{10,11}. Santosh et al. \cite{11} utilized lung symmetry in human chest radiography to detect pulmonary abnormalities. They extracted different features based on edge, texture and shape and combined Random Forest, Multi-Layer Perceptron (MLP) and Bayesian Network models to classify lungs into symmetric or asymmetric classes. Their study did not consider chest segmentation, and the chest segmentations of both hemithoraces were provided beforehand. Furthermore, they did not examine the varying exposure conditions that may result from poor image acquisition.

In the present study, we propose a novel method to segment hemithoraces by segmenting the ribs and the spine from canine and feline VD and DV radiographs. We then utilized the segmented ribs and fit an active contour to form the thorax region. The segmented spine was then used as a symmetry line to divide the segmented thorax into left and right hemithoraces contours. To compare left and right hemithoraces for making the final symmetry classification decision, we extracted seven features which captured shape and size from both sides. Finally, we utilized these features for training Support Vector Machines (SVM), Gradient Boost Classifier (GBC) \cite{12} and MLP models and made the final classification decision by majority voting.

To the best of our knowledge, this is the first research study proposing a method for segmenting hemithoraces and classifying symmetry in canine and feline radiography. We evaluated our approach on a dataset of 900 canine and feline radiographs from DV and VD views acquired at the Ontario Veterinary College (OVC) and labelled by an expert veterinary radiologist. Additionally, to prove that exposure setting or obstruction of the lungs does not affect the output of our model, we extensively compared the segmentation of the thorax in these scenarios with the ground-truth segmentation to study the drop in performance.

The main contributions of the present work are:
\begin{itemize}
\item Proposing a novel thorax segmentation method by segmenting the ribs and the spine and then fitting a snake to the rib region. This thorax segmentation method could be adopted for human chest radiography.
\item Proposing a method to classify symmetry in canine and feline radiography, which could also be used for image acquisition quality control.
\item Performing extensive analysis on the performance of the proposed method under different exposure settings to test its robustness to these extreme conditions and provide a detailed comparison.
\end{itemize}

\begin{figure}
    \centering
    \includegraphics[clip, trim=0.5cm 10cm 0.5cm 23cm, width=1\linewidth]{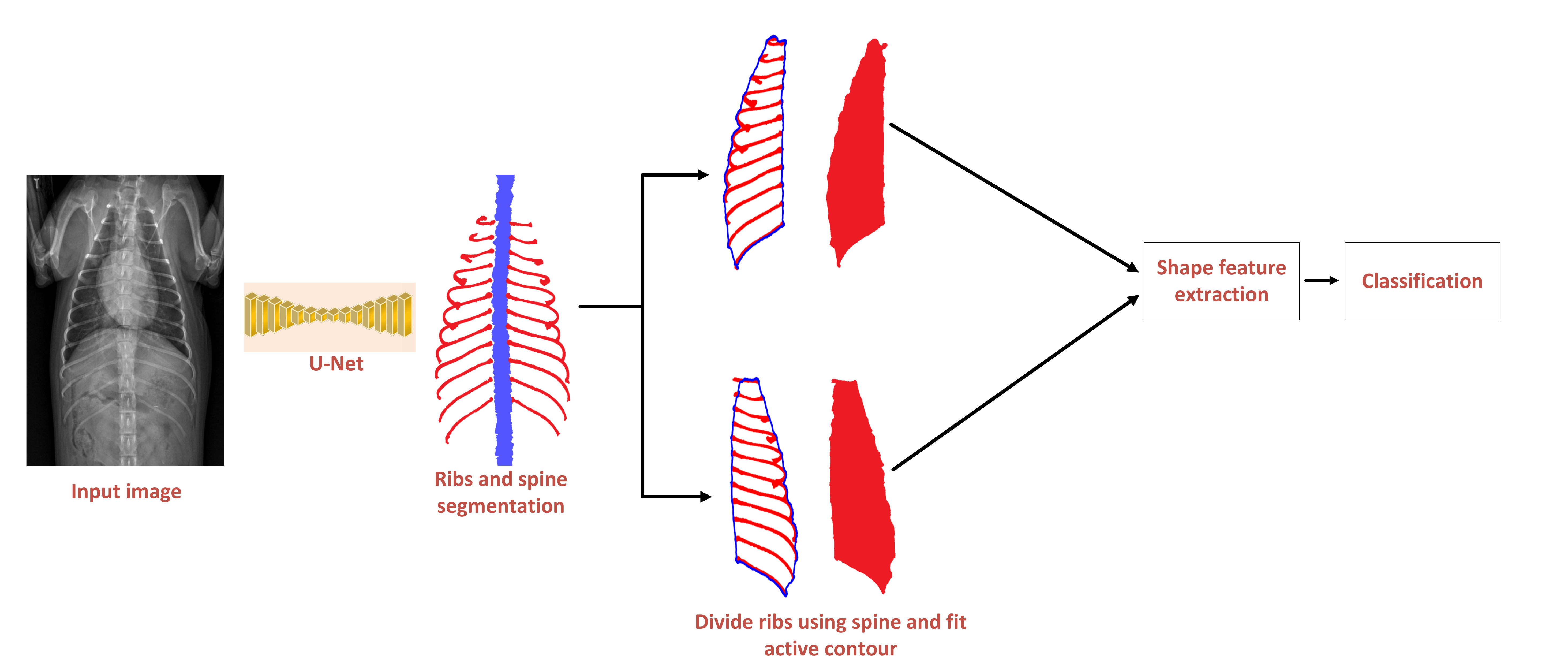}
    \caption{Automated symmetry assessment pipeline. First, a U-Net model masks the ribs and spine, then an active contour is fitted to the ribs to identify the thorax region. Afterwards, the thorax mask is divided into left and right regions, which form left and right hemithoraces. Finally, features are extracted from both hemithoraces for the final classification.
}
    \label{fig2}
\end{figure}

\section{Method}

The pipeline for our symmetry classification method is shown in Fig. \ref{fig2}. Our method automatically segments left and right hemithoraces and then uses them to make classification decisions. This section first explains the hemithoraces segmentation method, followed by feature extraction and classification. The proposed method for the segmentation of the ribs is not sensitive to the appearance of the lung in the radiograph and would function properly as long as the first sternal ribs and the last pair of floating ribs are presented on both left and right hemithoraces.

\subsection{Hemithoraces segmentation}

We started by segmenting the ribs and the spine using a U-Net \cite{13} model. This architecture consists of a series of convolutional blocks in an encoder-decoder fashion. Each encoder layer extracts features and downsizes the image by half until the bottleneck layer is obtained after this layer, the decoder stream or the upstream starts, where on each step, we upsampled the input features to finally reach an image with the same width and height as the input. The main point in the upstream section is the presence of skip connections \cite{9} to avoid vanishing gradients and improve gradient flow. Here, we concatenated features on the same level from the encoder and decoder before upsampling. Finally, U-Net generates a mask with the same width and height as the input image. We used pre-trained weights from ImageNet \cite{14} to increase training speed and faster convergence.

After segmenting the spine and ribs to detect left and right hemithoraces, we used active contours(deformable snakes) \cite{15} to find the thorax region. Since the segmented ribs can be considered as the approximate shape of the thorax boundary, we used the active contour method as they are designed to solve problems where the approximate shape of the boundary is known, and it preserves the topology. Specifically, the top-most rib, bottom-most rib, and the endpoints on the left and right of each rib approximately specify the thorax region for our method. An elastic snake is a contour represented parametrically as $v(s) = (x(s),y(s))$, where $x(s)$ and $y(s)$ are coordinates along the contour and $s\in[0,1]$, which will be used to minimize an energy function consisting of internal energy $(E_{internal})$ and external energy $(E_{external})$. The internal energy generally controls the shape of the snake itself and consists of elasticity and stiffness as described in below equation:

\begin{equation}
    E_{internal}(v(s)) = \alpha(s)|\frac{dv}{ds}|^2 + \beta(s)|\frac{d^2v}{ds^2}|^2
\end{equation}

where the elasticity term$|\frac{dv}{ds}|^2$  controls the length of the snake, and stiffness term $|\frac{d^2v}{ds^2}|^2$  controls its curvature or how much it is allowed to bend to fit existing boundaries. External energy describes how well the curve matches the image data locally. A simple external energy function was described based on a first-order gradient of the image $I$ as:
\begin{equation}
    E_{external}(v(s)) = -|\nabla I(x,y)|^2
\end{equation}
The total energy of the snake is the sum of its external and internal energy:
\begin{equation}
    E_{snake} = \int _{0}^{1}E_{internal} (v(s)) + E_{external} (v(s)) ds
\end{equation}

To optimize this function using gradient descent, we needed an initial guess. In this study, we used a rectangle as the initial shape, which is minimal and covers all four corners of the segmented ribs outputted by the U-Net model. After segmenting the thorax, we used the spine as the symmetry line to divide the thorax mask into left and right regions. This step is represented in Fig. \ref{fig2}. 

\subsection{Shape feature extraction}

After segmenting left and right hemithoraces separately, we focused on extracting distinctive features from both contours to be used in an ensemble of SVM, GBC and MLP. The low-level shape features \cite{16} and shape histograms that were extracted include:
\begin{itemize}
    \item Area: For a segmented hemithorax region $s\in R^2$, $s=(x_i,y_i)_{i=1,...,N}$, it returns a scalar specifying the number of pixels inside the segmented hemithorax.
\item Perimeter: For a segmented hemithorax region $s\in R^2$, $s=(x_i,y_i)_{i=1,...,N}$, it returns the number of pixels over the region's boundary.
\item Centroid horizontal distance ($c_{xo}$): The center of mass of the segmented hemithorax $s\in R^2$ $s=(x_i,y_i)_{i=1,...,N}$ or its centroid $c_o=(x_o,y_o)$ can simply be expressed as $x_o=\sum_{i=0}^N\frac{x_i}{N}$, $y_o=\sum_{i=0}^N\frac{y_i}{N}$. To compare the centroid horizontal distance of the left and right hemithoraces, we found the distance on the x-axis from each centroid to the closest point on the spine.
\item First-rib-width ($w_x$): This feature compares the distance of the first rib endpoints on the left and right hemithoraces to the spine midline in pixels. As shown in Fig. \ref{fig3}, to calculate this value, we found the top left point on the left hemithorax and the top right point on the right hemithorax, which represents the end point of the first left ribs and first right rib, respectively, and calculated their euclidean distance to the spine midline. Red arrows show this distance on both the right and left sides in Fig. \ref{fig3}, and green dotted circles delineate the endpoints.

\begin{figure}
    \centering
    
    \begin{subfigure}[b]{0.45\textwidth}
        \includegraphics[width=1.0\linewidth,height=3.05in]{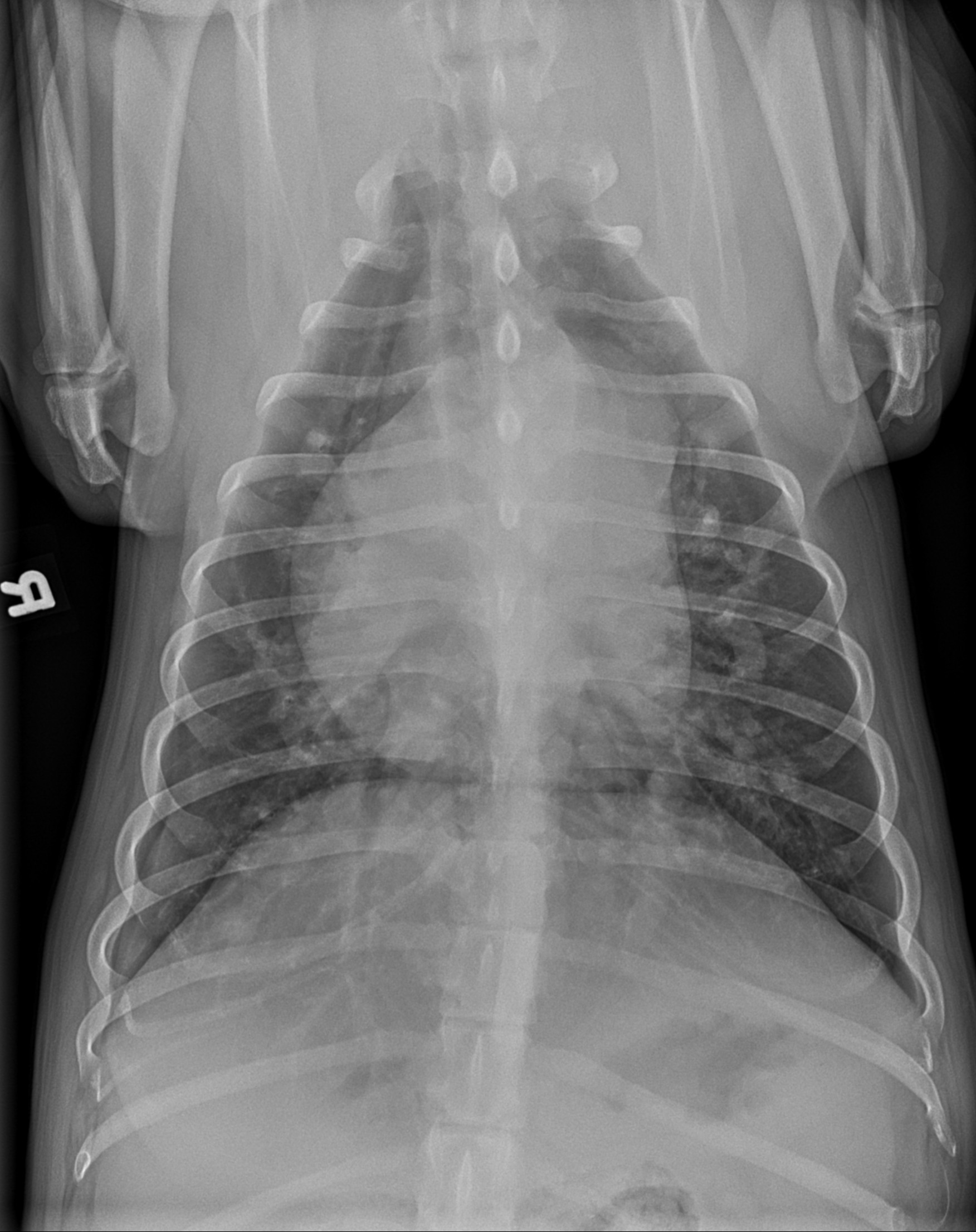}
    \end{subfigure}
    \begin{subfigure}[b]{0.45\linewidth}
        \includegraphics[width=1.0\linewidth]{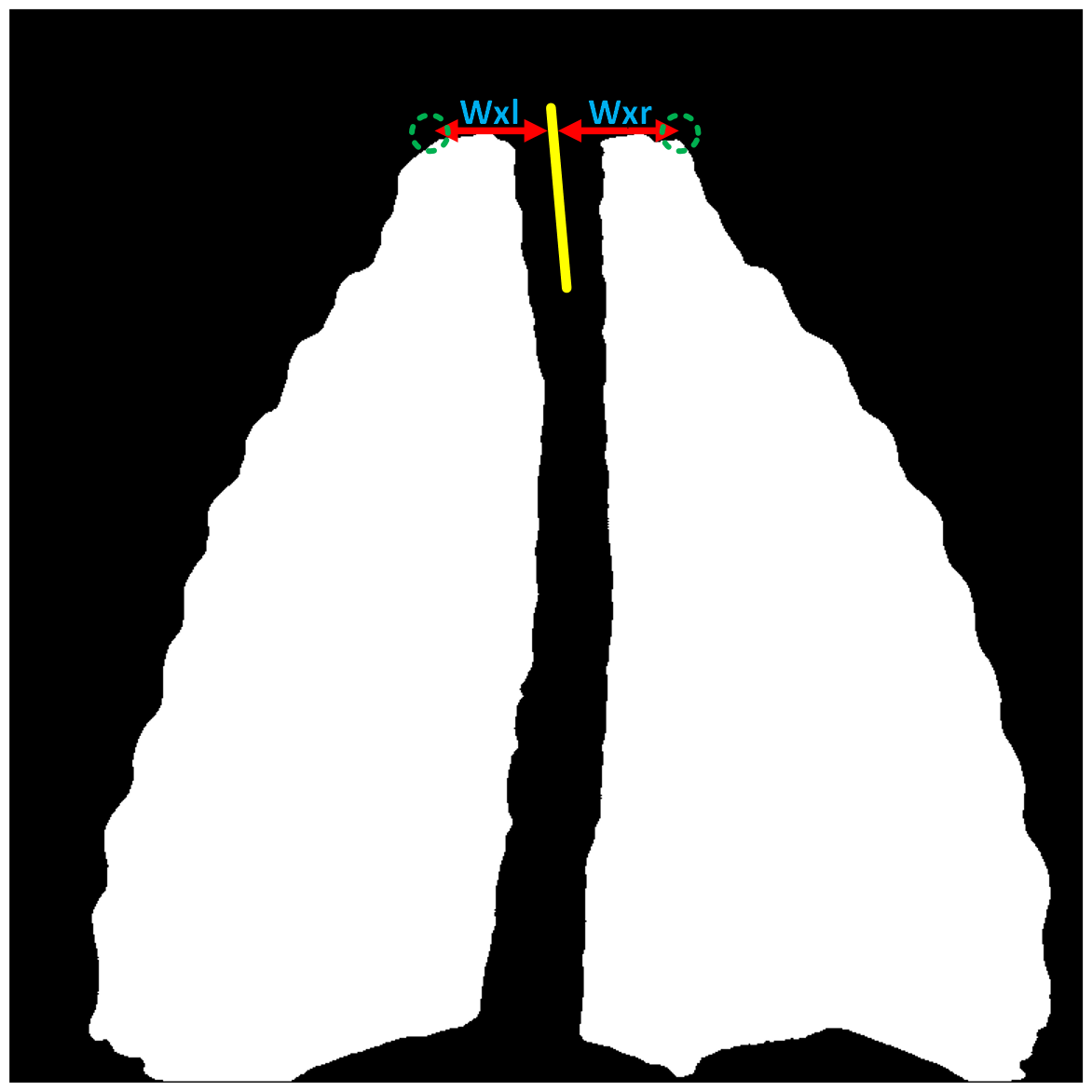}
    \end{subfigure}
    
    \caption{The first-rib-width ($w_x$) index is calculated by measuring the distance in pixels, along the x-axis, between the spine midline (shown by the yellow line) and the endpoints of first ribs on left and right hemithoraces (shown by green dotted circles). Subscripts $r$ and $l$ represent the feature measured in the right and left hemithoraces, respectively. }
    \label{fig3}
\end{figure}

\item Contour density plot. Suppose ${X} = \{x(i,j) | 1 \leq i \leq h, 1\leq j \leq w\}$ of size $w\times h$ denote the image containing segmented left or right hemithorax region  where $x(i,j)$ denotes the graylevel value at location $(i,j)$, we can calculate the horizontal and vertical histogram of the image using below equation:
\begin{equation}
\label{eq4}
\begin{split}
    H_j = \sum_i^wx(i,j)
    \\
    W_i = \sum_j^wx(i,j)
    \end{split}
\end{equation}

After calculating vertical and horizontal histograms of both left and right hemithoraces using the above equation, we used Jenson-Shannon Divergence (JSD) and intersection to compare the vertical and horizontal histograms between the left and right regions. The JSD for two probability distributions of $H_1$ and $H_2$ is defined as
\begin{equation}
\label{eq5}
\centering
    \begin{split}
        JSD(H_1,H_2) = \frac{1}{2} KL(H_1, M) + \frac{1}{2}KL(H_2,M)\\
        where \ M=\frac{1}{2}(H_1+H_2)\\
        KL(H_1, H_2) = \sum_XH_1(X)\log\frac{H_1(X)}{H_2(X)}
    \end{split}
\end{equation}

where KL represents kullback-leibler divergence. The intersection between two histograms can be defined as
\begin{equation}
    d(H_1, H_2) = \sum_Xmin(H_1(X), H_2(X))
    \label{eq6}
\end{equation}

\begin{figure}[h]
    \centering
    \includegraphics[width=0.5\linewidth, height=3.5in]{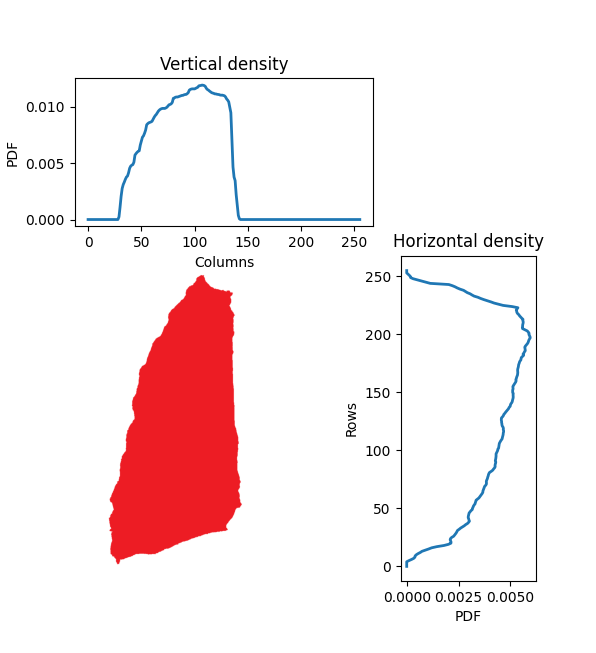}
    \caption{Horizontal and vertical histograms of a hemithorax. We compute this histogram for both left and right hemithoraces and then find their similarity using JSD and intersection as defined in equation \ref{eq5} and \ref{eq6}.}
    \label{fig4}
\end{figure}

\item Intersection over Union (IoU). In order to measure how two contours match each other, we could find their overlap. However, we could not simply flip the left hemithorax and find its IoU with the right one since hemithoraces do not always happen in the center of the radiograph. Henceforth, we registered the left hemithorax to the right hemithorax using translation and then we found IoU using the below equation:

\begin{equation}
    IoU = \frac{left\cap right}{left \cup right}
\end{equation}
\end{itemize}

To quantify symmetry, we used a similarity index $sim(r_L, r_R)$ for 6 extracted features from left and right hemithoraces, where  $r_L$ and $r_R$ are features in the left and right hemithoraces, respectively. We did not use the similarity index for the IoU feature since it is already in the range of $[0,1]$.
For low-level shape features such as area, perimeter, centroid horizontal distance and first-rib -width, we defined the similarity index as below, which is a ratio in the range of $[0,1]$. The similarity index of 1 and 0 represents pure symmetry and asymmetry, respectively.
\begin{equation}
    sim(r_L, r_R) = \frac{min(r_L, r_R)}{max(r_L, r_R)}
\end{equation}

\subsection{Classification of symmetry}

After encoding the radiograph with seven features and calculating their corresponding similarity index, we used an ensemble of three classifiers, SVM, MLP and GBC, for canine and feline radiograph symmetry classification.

SVM aims to find a hyperplane in an N-dimensional space that distinctly classifies the data points. It transforms data using a kernel function to a space where data can be separated using a line. The choice of the kernel is of utmost importance because it has a direct impact on the performance of the model. Common kernels for training an SVM model include Radial Basis Functions (RBF) and polynomial functions of different degrees. To find the best kernel, we used five-fold cross-validation.

MLP uses a series of connected neurons to pass information in a feed-forward fashion to make classification decisions. To tune the MLP model for the best classification decision, we then utilized errors made on each sample and updated each neuron's weight and bias using an optimization technique such as Gradient Descent. The MLP has several hyperparameters to tune, such as learning rate, regularization term, number of hidden units, network depth, and more. In the present study, we used five-fold cross-validation to find the best set of hyperparameters.

GBC uses the same dataset to generate many decision trees, each to decrease previous errors. At each step, decision trees were built using the errors our model had made thus far. The final decision for a gradient boost classifier is the sum of the decisions of all the decision trees it generated, each multiplied by a factor called the learning rate. In this model, the learning rate, maximum number of decision trees and maximum depth in each decision tree are among the many hyperparameters that were tuned by five-fold cross-validation. 

We used an ensemble of the three above-mentioned methods to make the final symmetry classification decision. Each model's hyperparameters were tuned independently and then merged to form the final ensemble model.
\section{Experiments}
\subsection{Dataset}

Since there was no public dataset available for this task, a dataset of 900 VD and DV radiographs of dogs and cats was retrieved from the OVC database.  Radiographs were acquired from various computed radiography and direct digital radiography systems at referral practices (primary care clinics) and at OVC. Images were provided to the OVC in either JPEG or DICOM format and stored in the Picture Archiving and Communication System (PACS) provided by AGFA \cite{17}, with the patient file and any onsite imaging. To gather the dataset of this study, radiographs were exported as JPEG and anonymized either by removal of patient data from the images prior to exportation or through blurring data burned into the image as was present in many images from referral practices. Among 900 radiographs, 270 represented an asymmetric thorax, while the rest of them contained a symmetric thorax. Note that for training and evaluating our model, we did not exclude overexposed or underexposed radiographs.

\subsection{Implementation details}

Our proposed method was implemented in Python, and our deep learning models were implemented using the Pytroch library \cite{18}. The model was trained on an NVIDIA RTX3090 24GB GPU and an AMD Ryzen 9 5950X CPU @ 3.4 GHz $\times 16$. All models were trained with a batch size of 2 and resized to size 1024 x 1024. We used the AdamW optimizer \cite{20} with an initial learning rate of 0.0001. We used early stopping to avoid overfitting. Our model was evaluated using 5-fold cross-validation with a fixed split across all experiments. We initialized our model weights with a pre-trained model on ImageNet \cite{14}. Considering hyperparameters, for training segmentation model loss, we used focal loss\cite{21} with $\alpha= 0.8$ and Tversky loss \cite{19} with $\alpha=2.0$ and $\beta = 2.0$. For the MLP symmetry classification model, we used 50 neurons in three hidden layers, adam solver, and ReLu as the activation function. For the GBC model, we used 100 estimators, each with a maximum depth of 3 and a learning rate of 0.1. For the SVM symmetry classification model, we used an RBF kernel with $\gamma = 0.01$ and $C=0.1$.

\subsection{Experimental design}

This section describes the performance evaluation of the proposed thorax segmentation method under poor exposure settings or when part of the lung is completely obscured or missing. We applied a set of transformations on test radiographs on each fold and performed three tests: underexposure test, overexposure test and obstruction test.
The exposure settings of properly exposed radiographs (as labelled by an expert veterinary radiologist) were synthetically altered by performing a gamma transformation using the equation below, where $\gamma$  was sampled using a uniform distribution in the range of $[0.2 , 0.5]$ and $[2.0 , 5.0]$ to simulate underexposed and overexposed  radiographs, respectively. 
\begin{equation}
    I(x,y) = I(x,y)^\gamma
\end{equation}

\begin{figure}[h]
    \centering
    \begin{subfigure}[b]{0.24\textwidth}
        \includegraphics[width=1.0\linewidth,height=2.05in]{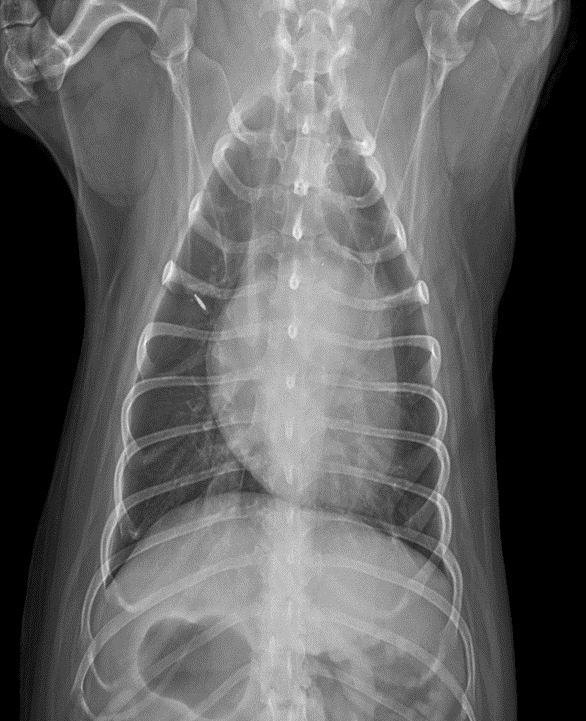}
    \end{subfigure}
    \begin{subfigure}[b]{0.24\textwidth}
        \includegraphics[width=1.0\linewidth,height=2.05in]{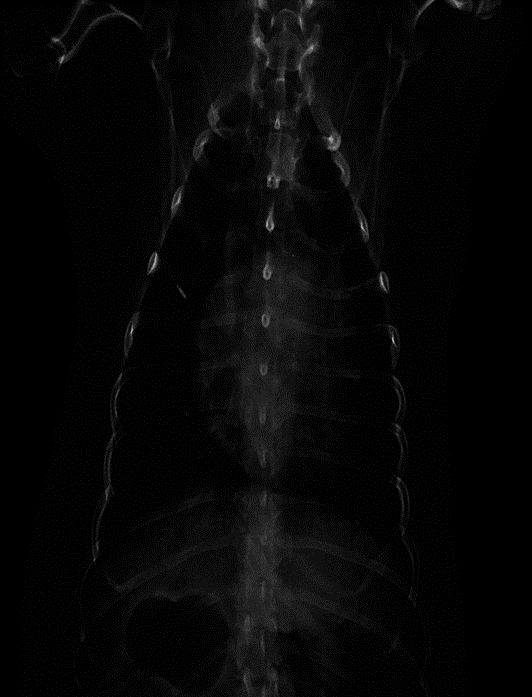}
    \end{subfigure}
    \begin{subfigure}[b]{0.24\textwidth}
        \includegraphics[width=1.0\linewidth,height=2.05in]{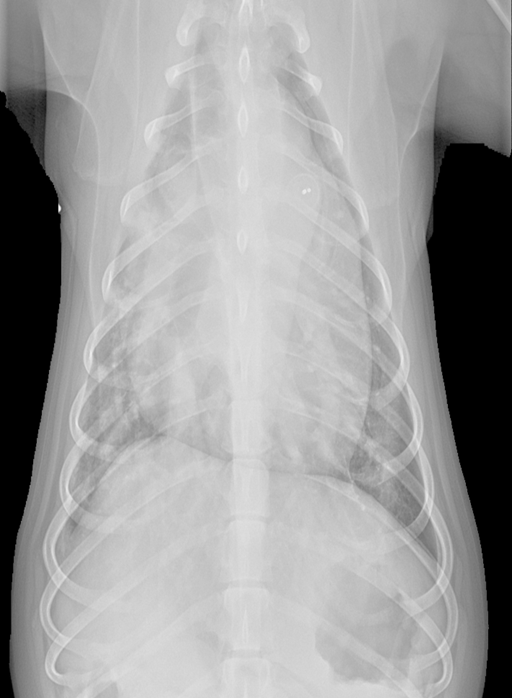}
    \end{subfigure}
    \begin{subfigure}[b]{0.24\textwidth}
        \includegraphics[width=1.0\linewidth,height=2.05in]{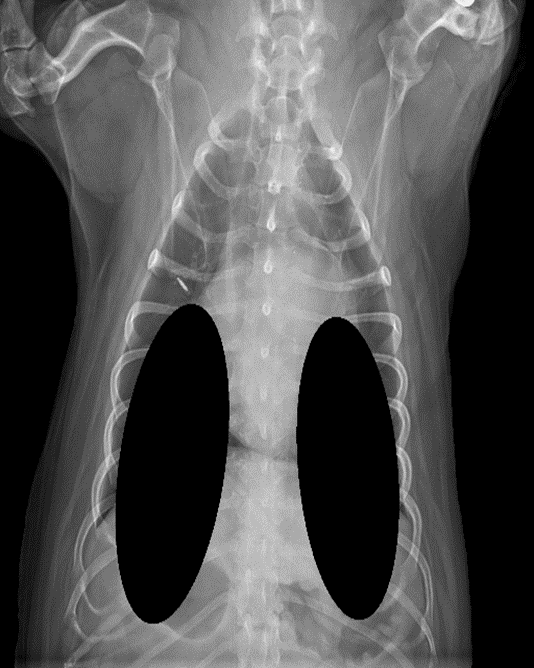}
    \end{subfigure}
    \vfill
    \begin{subfigure}[b]{0.24\textwidth}
        \includegraphics[width=1.0\linewidth,height=2.05in]{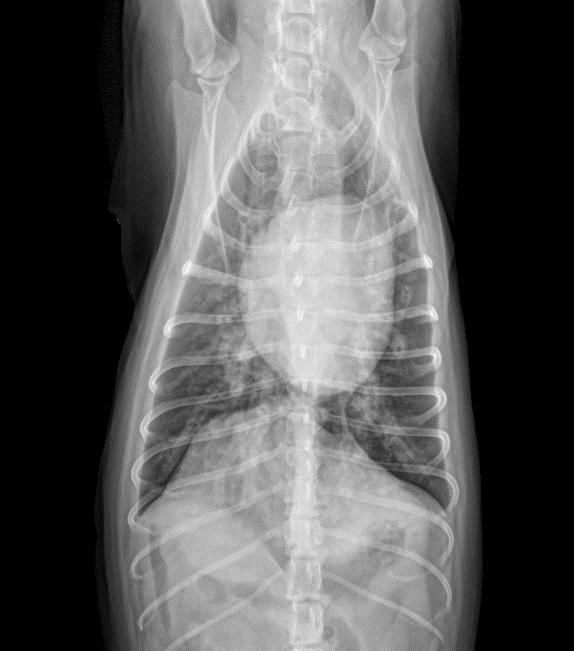}
        \caption{Input}
    \end{subfigure}
    \begin{subfigure}[b]{0.24\textwidth}
        \includegraphics[width=1.0\linewidth,height=2.05in]{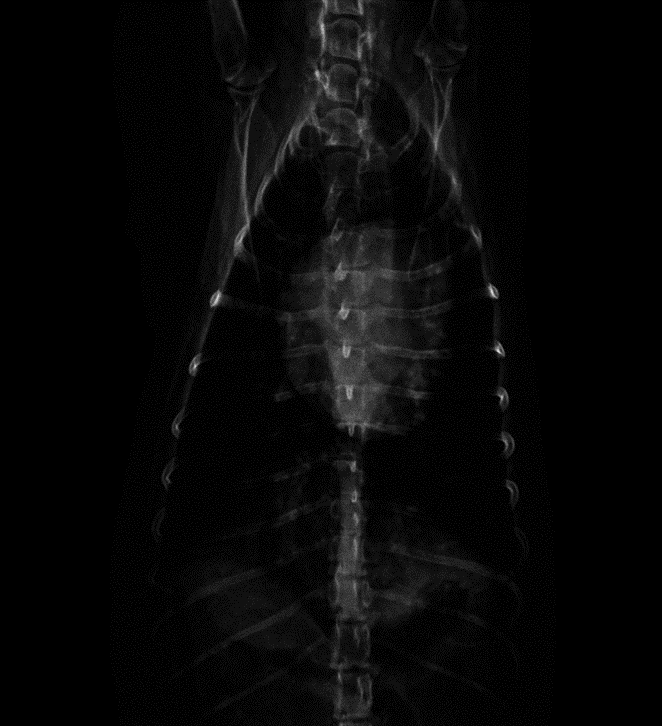}
        \caption{Overexposed}
    \end{subfigure}
    \begin{subfigure}[b]{0.24\textwidth}
        \includegraphics[width=1.0\linewidth,height=2.05in]{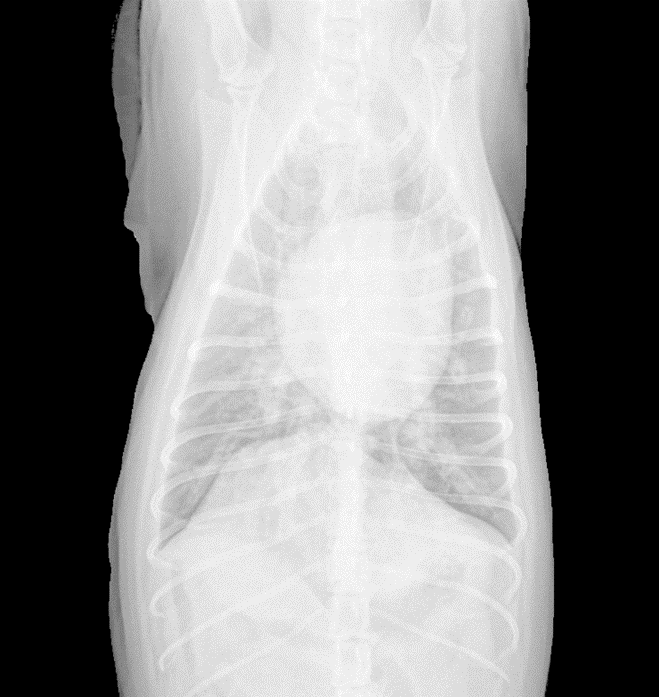}
        \caption{Underexposed}
    \end{subfigure}
    \begin{subfigure}[b]{0.24\textwidth}
        \includegraphics[width=1.0\linewidth,height=2.05in]{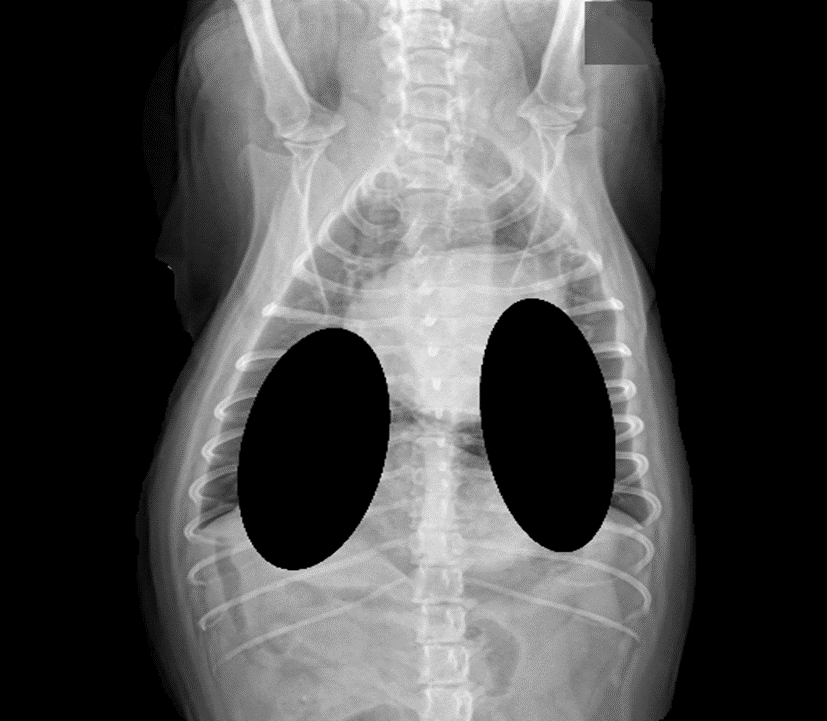}
        \caption{Obscured}
    \end{subfigure}
    \caption{Examples of two normal radiographs with their corresponding synthetic overexposed, underexposed, and obscured transformation.}
    \label{fig5}
\end{figure}

After applying the gamma transformation, we added gaussian noise with  $\mu= 0$ and $\sigma \in [0, 10]$. The results following the application of these transformations on two different radiographs are shown in Fig. \ref{fig5}.
To obscure the lungs for the obstruction test, we used ground-truth segmentation for the thorax achieved by fitting an active contour to the ground-truth ribs and fitted an ellipse to it. Then, we randomly divided major and minor axis lengths by values in the range $[2,4]$ and overlapped them with the original image to simulate obscured lungs. Figure \ref{fig5} shows the results of applying this process to two different radiographs. Note that, for this experiment, we only applied these transformations to the test set of each fold.

To evaluate the performance of the proposed method, we applied a direct segmentation method for segmenting hemithoraces and compared its outcome with the proposed method using the IoU metric between the ground-truth left and right hemithoraces and the models’ prediction. 
In the direct segmentation approach, we directly utilized hemithoraces masks, which we obtained by fitting active contour to ground-truth rib labels, to train a deep learning model that directly outputs both hemithoraces by feeding a radiograph as an input. The U-Net model and the same training scheme with slight modifications were used for the direct segmentation method.

\section{Results}
\subsection{Thorax segmentation}

\begin{figure}
    \centering
    
    \begin{subfigure}[b]{0.24\textwidth}
        \includegraphics[width=1.0\linewidth,height=2.05in]{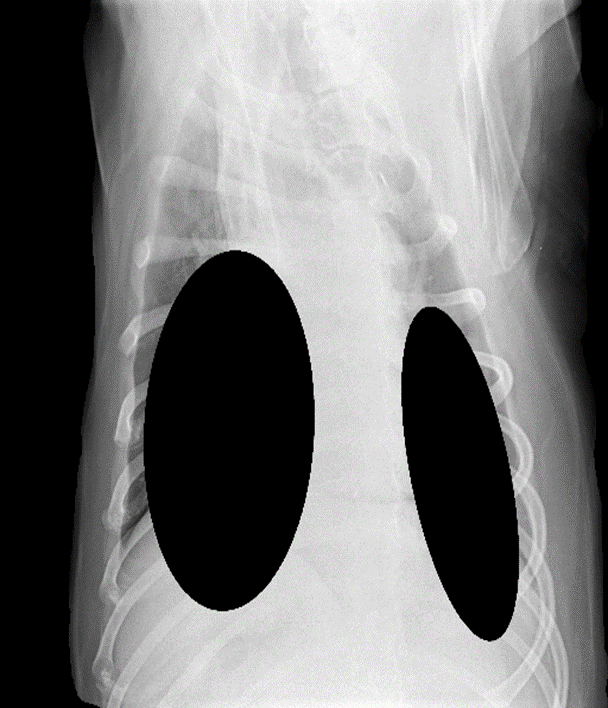}
    \end{subfigure}
    \begin{subfigure}[b]{0.24\textwidth}
        \includegraphics[width=1.0\linewidth,height=2.05in]{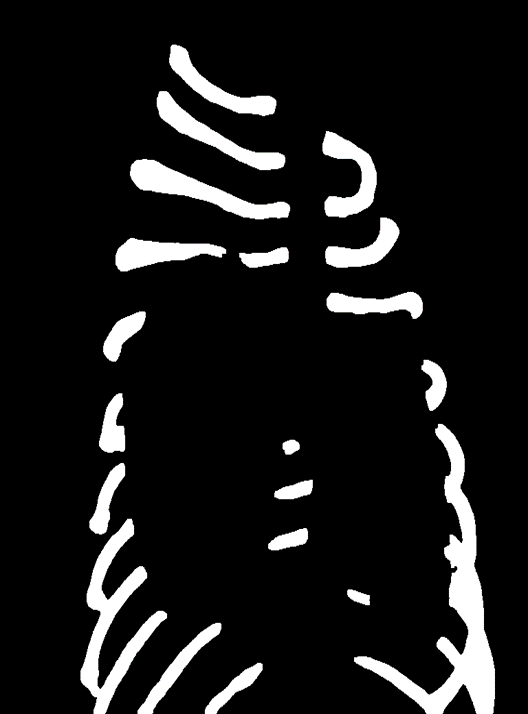}
    \end{subfigure}
    \begin{subfigure}[b]{0.24\textwidth}
        \includegraphics[width=1.0\linewidth,height=2.05in]{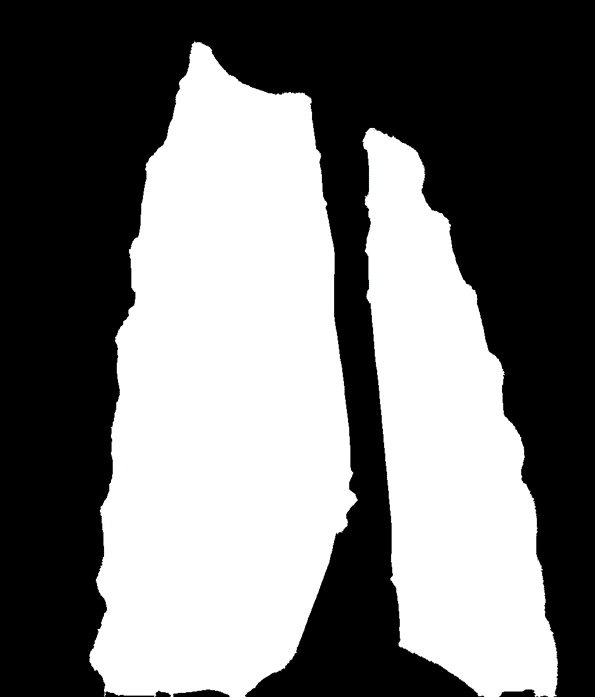}
    \end{subfigure}
    \begin{subfigure}[b]{0.24\textwidth}
        \includegraphics[width=1.0\linewidth,height=2.05in]{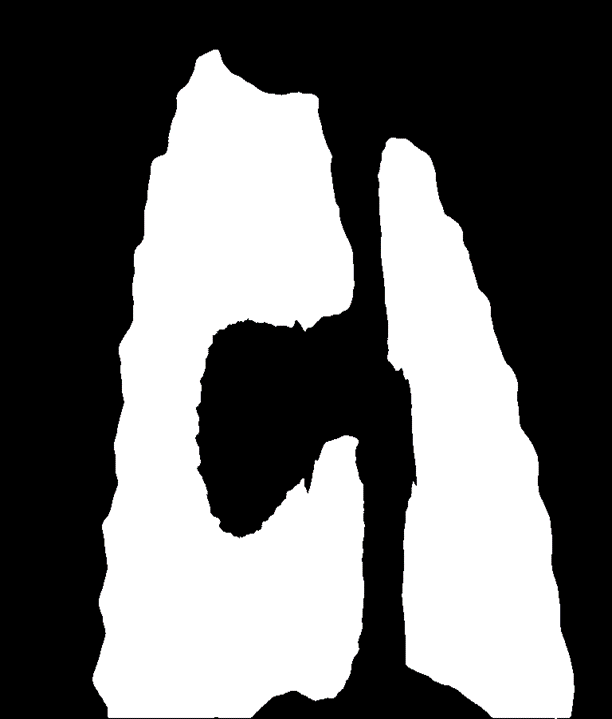}
    \end{subfigure}
    \vfill
    \begin{subfigure}[b]{0.24\textwidth}
        \includegraphics[width=1.0\linewidth,height=2.05in]{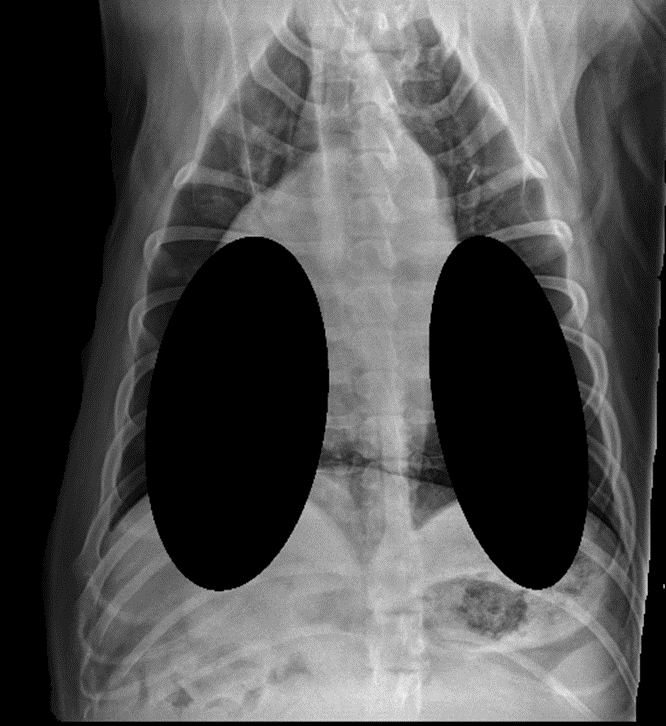}
    \end{subfigure}
    \begin{subfigure}[b]{0.24\textwidth}
        \includegraphics[width=1.0\linewidth,height=2.05in]{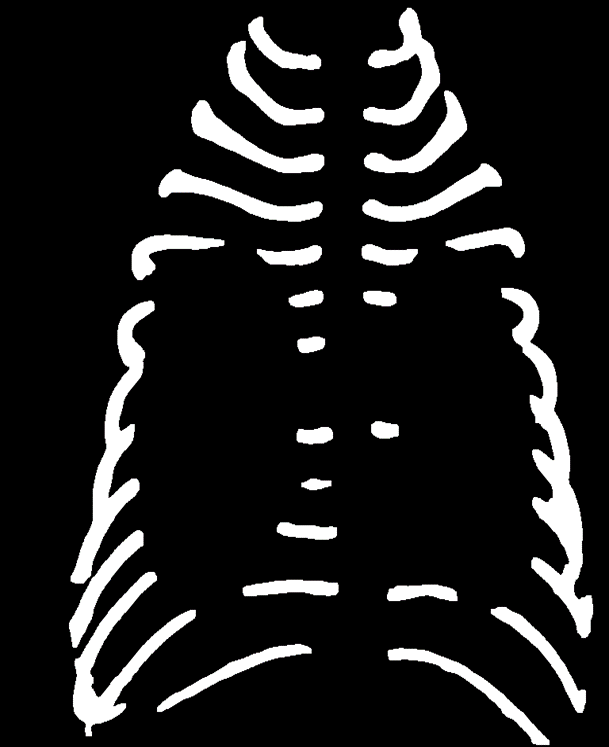}
    \end{subfigure}
    \begin{subfigure}[b]{0.24\textwidth}
        \includegraphics[width=1.0\linewidth,height=2.05in]{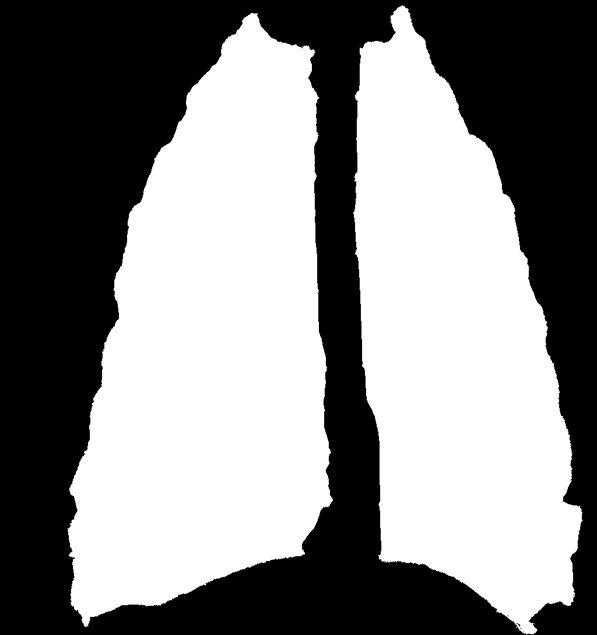}
    \end{subfigure}
    \begin{subfigure}[b]{0.24\textwidth}
        \includegraphics[width=1.0\linewidth,height=2.05in]{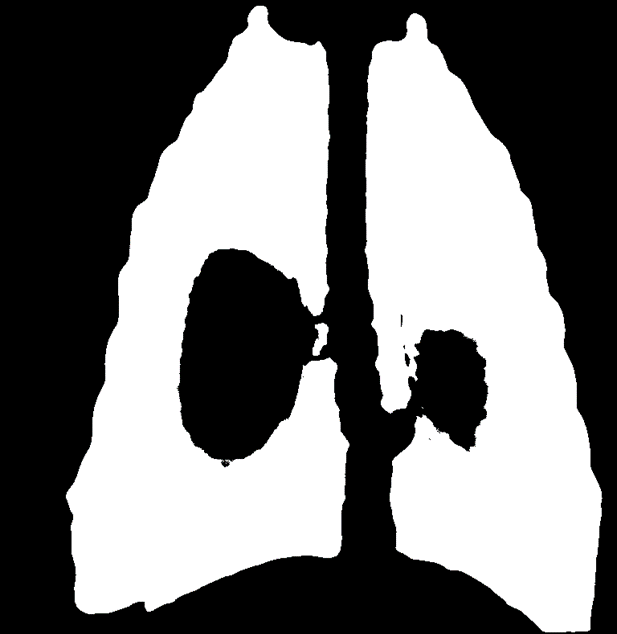}
    \end{subfigure}
    \vfill
    \begin{subfigure}[b]{0.24\textwidth}
        \includegraphics[width=1.0\linewidth,height=2.05in]{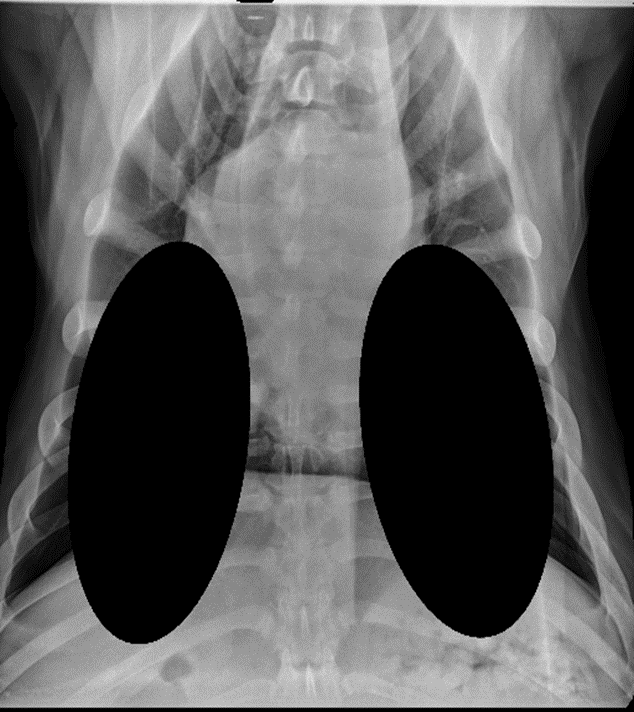}
        \caption{Input}
    \end{subfigure}
    \begin{subfigure}[b]{0.24\textwidth}
        \includegraphics[width=1.0\linewidth,height=2.05in]{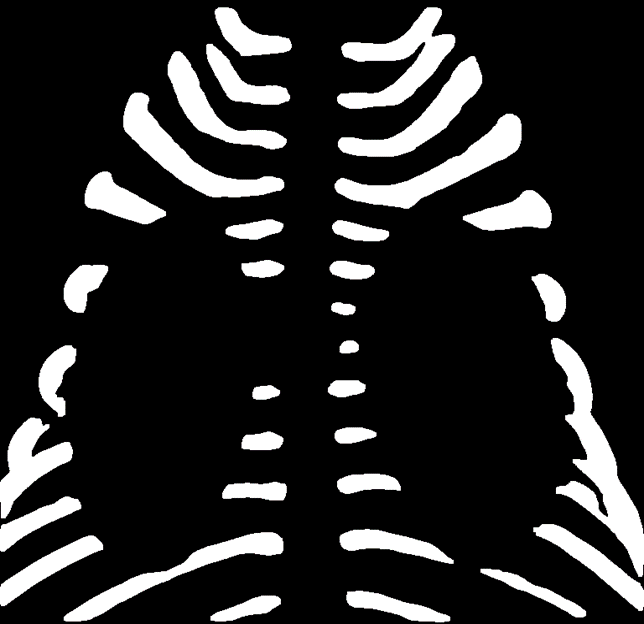}
        \caption{Obscured ribs}
    \end{subfigure}
    \begin{subfigure}[b]{0.24\textwidth}
        \includegraphics[width=1.0\linewidth,height=2.05in]{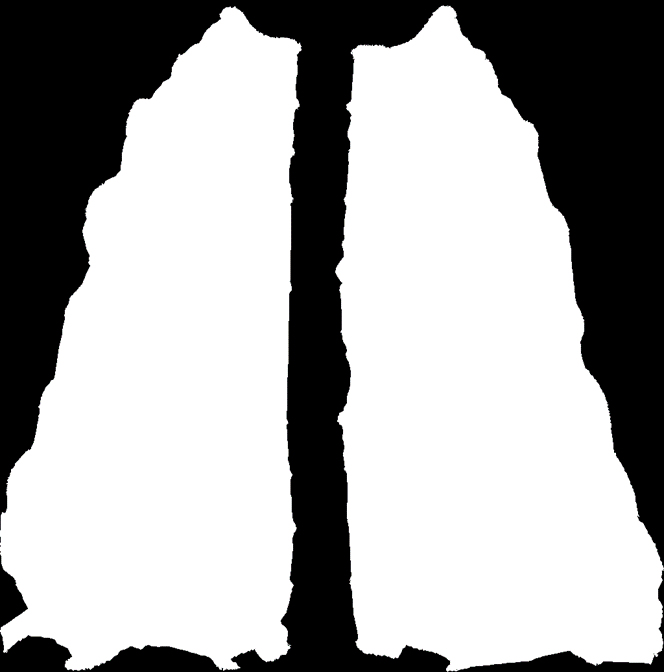}
        \caption{Proposed method}
    \end{subfigure}
    \begin{subfigure}[b]{0.24\textwidth}
        \includegraphics[width=1.0\linewidth,height=2.05in]{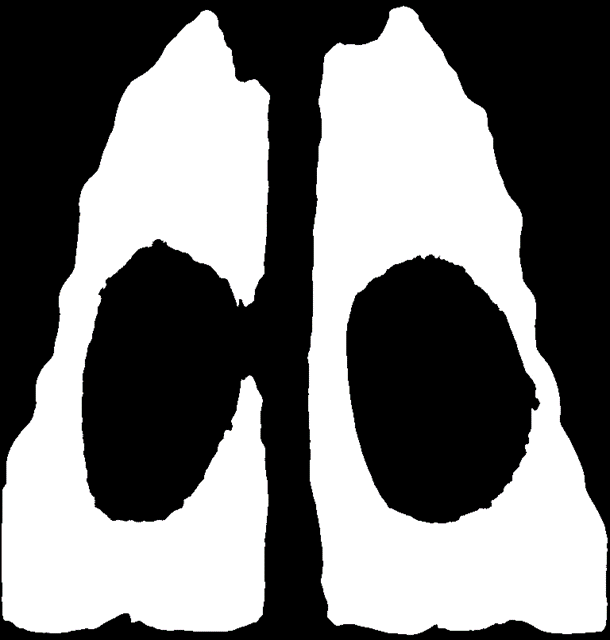}
        \caption{Direct segmentation}
    \end{subfigure}

    \caption{Comparison between the hemithoraces segmentation outputs of the proposed model and a conventional direct segmentation method on 3 sample radiographs obstructed synthetically.}
    \label{fig6}
\end{figure}

A side-by-side comparison of the prediction of the proposed model and the direct segmentation method is provided in Fig. \ref{fig6}. In the first row, the radiograph is underexposed and obscured. Even though part of the ribs is completely obstructed, our model could still recover the entire shape of hemithoraces, while the direct segmentation method failed to recover the whole shape. This is because the active contour model aligned the shape to the edges of the image and was less sensitive to the object's internal structure. The same pattern also happened for images in the second and third rows. The proposed model effectively segmented the hemithoraces when the boundary of the ribs is visible in the radiograph. 

As can be seen in Table \ref{tb1}, our proposed method achieved a slightly higher IoU compared to the direct segmentation method when using normal radiographs. Note that here, The term normal refers to radiographs within the test that are not overexposed or underexposed as labelled by a radiology expert. However, the IoU difference between the proposed and direct segmentation methods was higher with 2.04$\%$ and $1.16\%$ when we used the underexposed and overexposed radiographs, respectively. The most significant difference between our proposed and direct segmentation method occurred when we obscured part of the hemithoraces with a difference of $6.1\%$. Additionally, the IoU of our model only dropped by $3.31\%$ when we obstructed lungs, while the IoU of the direct segmentation method dropped by $8.35\%$  after introducing the obstruction (Table \ref{tb1}). Overall, our proposed method worked better than the direct segmentation approach, mainly when the radiograph provided was either overexposed or underexposed.

\begin{table}[h]
\caption{Comparison between the conventional direct segmentation and proposed segmentation methods using normal, underexposed, overexposed, and obscured radiographs.}
\centering
\begin{tabular}{|c|cc|c|}
\hline
\multirow{2}{*}{Radiograph type} & \multicolumn{2}{c|}{IoU}                                  & \multirow{2}{*}{Difference} \\ \cline{2-3}
                                 & \multicolumn{1}{c|}{Proposed model} & Direct segmentation &                             \\ \hline
Normal                           & \multicolumn{1}{c|}{94.72}          & 93.66               & 1.06                        \\ \hline
Underexposed                     & \multicolumn{1}{c|}{91.89}          & 89.85               & 2.04                        \\ \hline
Overexposed                      & \multicolumn{1}{c|}{92.57}          & 91.41               & 1.16                        \\ \hline
Obscured                         & \multicolumn{1}{c|}{91.41}          & 85.31               & 6.1                         \\ \hline
\end{tabular}
\label{tb1}
\end{table}

\subsection{Symmetry detection}

\begin{figure}[h]
    \centering
    \includegraphics[width=0.5\linewidth, height=3in]{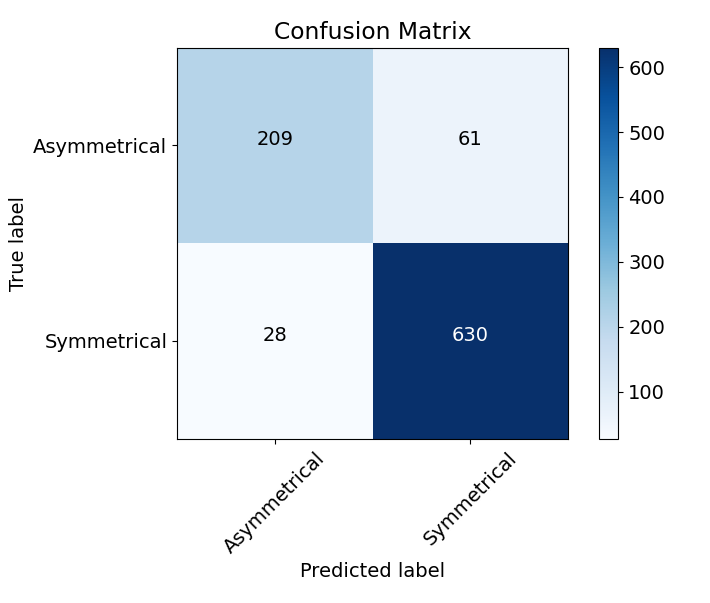}
    \caption{Confusion matrix for ensemble model}
    \label{fig7}
\end{figure}

Table \ref{tb2} provides the SVM, GBC and MLP model results across five-folds, with MLP resulting in the best performance. Employing an ensemble of three models and majority voting increases precision, recall, F1, and Area Under Curve (AUC) compared to base model values (Table \ref{tb2}). Additionally, the confusion matrix for the ensemble model is provided in Fig. \ref{fig7}. As can be seen, across all test examples in all  folds, our model predicted 209 asymmetric cases correctly out of 270 available cases. Additionally, among all 237 predicted cases as asymmetric, only 28 radiographs were incorrectly classified.
\begin{table}[h]
\caption{Results of symmetry classification using the SVM, MLP, gradient boost, and ensemble models.}
\centering
\begin{tabular}{|ll|l|l|l|l|}
\hline
\multicolumn{2}{|l|}{\diagbox[width=15em]{Model}{Metric}}               & Precision      & Recall         & F1             & Area Under Curve (AUC) \\ \hline
\multicolumn{2}{|l|}{SVM}                                       & 81.36          & 74.98          & 78.03          & 80.25                  \\ \hline
\multicolumn{2}{|l|}{MLP}                                       & \textbf{85.41} & \textbf{76.33} & \textbf{80.61} & \textbf{84.37}         \\ \hline
\multicolumn{2}{|l|}{GBC}                            & 82.19          & 75.65          & 78.78          & 83.12                  \\ \hline
\multicolumn{1}{|l|}{\multirow{6}{*}{Ensemble Model}} & Fold 1  & 96.66          & 86.66          & 91.22          & 93.32                  \\ \cline{2-6} 
\multicolumn{1}{|l|}{}                                & Fold 2  & 83.21          & 75.85          & 80.47          & 82.66                  \\ \cline{2-6} 
\multicolumn{1}{|l|}{}                                & Fold 3   & 84.61          & 72.25          & 77.94          & 81.54                  \\ \cline{2-6} 
\multicolumn{1}{|l|}{}                                & Fold 4   & 89.18          & 75.00          & 81.48          & 85.64                  \\ \cline{2-6} 
\multicolumn{1}{|l|}{}                                & Fold 5  & 87.87          & 78.63          & 82.99          & 83.34                  \\ \cline{2-6} 
\multicolumn{1}{|l|}{}                                & Overall & \textbf{88.30} & \textbf{77.67} & \textbf{82.82} & \textbf{85.3}          \\ \hline
\end{tabular}
\label{tb2}
\end{table}

The proposed active contour method could be applied in two manners to mask the right and left hemithoraces. The first approach (one-snake) includes applying the active contour to the entire segmented ribs to create the thorax region and then dividing the masked thorax using the spine method. The second method (two-snakes) is to first divide the segmented ribs into the left and right partitions using the spine and then fit an active contour to each side separately to mask the left and right hemithoraces regions. We compared the classification result of these two approaches in Table \ref{tb3}, where the one-snake performed better than the two-snakes in both F1 and AUC metrics, in addition to computational advantage. In fact, in the two-snakes approach, two active contour models should be applied, one on the left and one on the right rib partitions, while in the one-snake method, we only need to apply the active contour once.

\begin{figure}
    \centering
    
    \begin{subfigure}[b]{0.19\textwidth}
        \includegraphics[width=1.0\linewidth,height=1.2in]{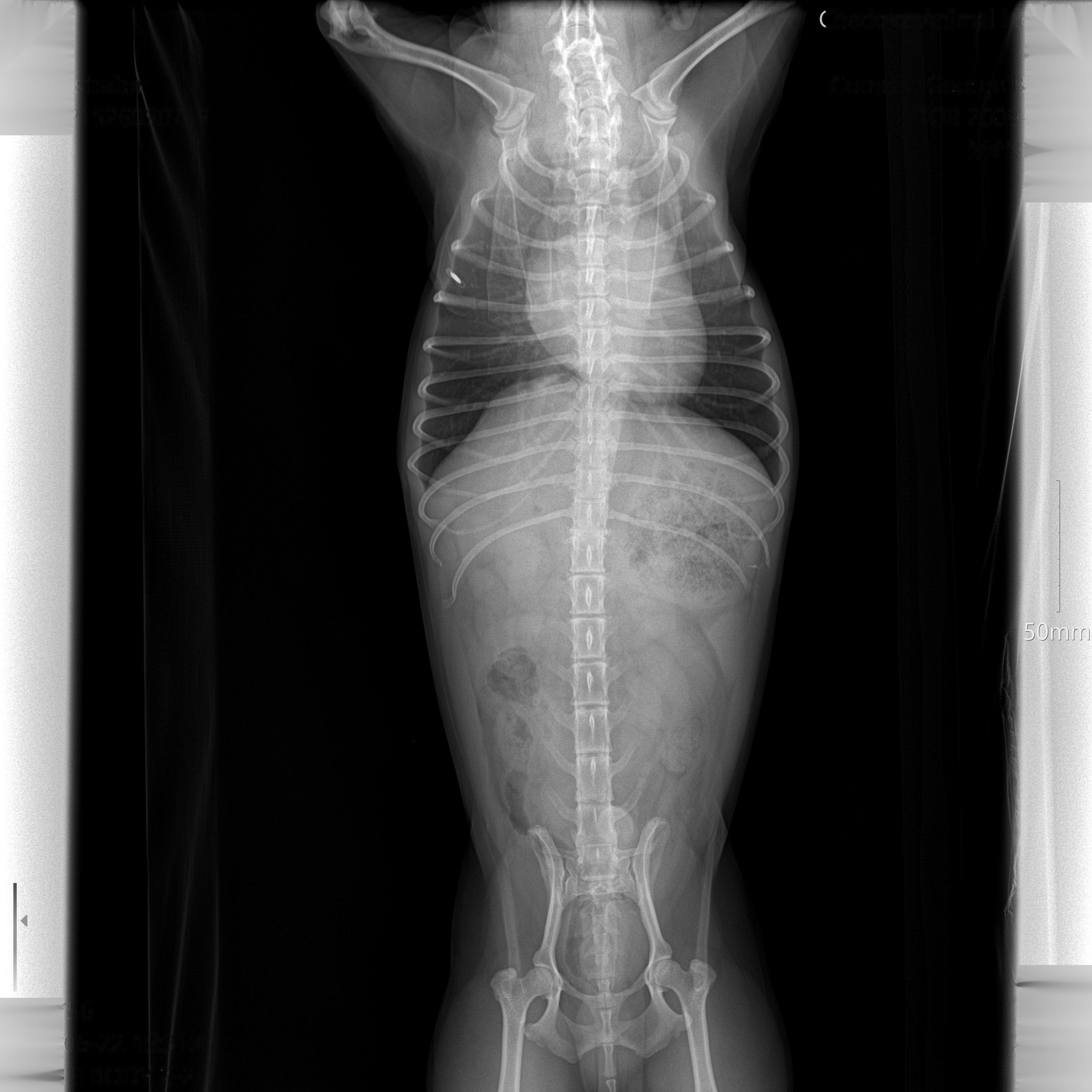}
        \caption{Input}
    \end{subfigure}
    \begin{subfigure}[b]{0.19\textwidth}
        \includegraphics[width=1.0\linewidth,height=1.2in]{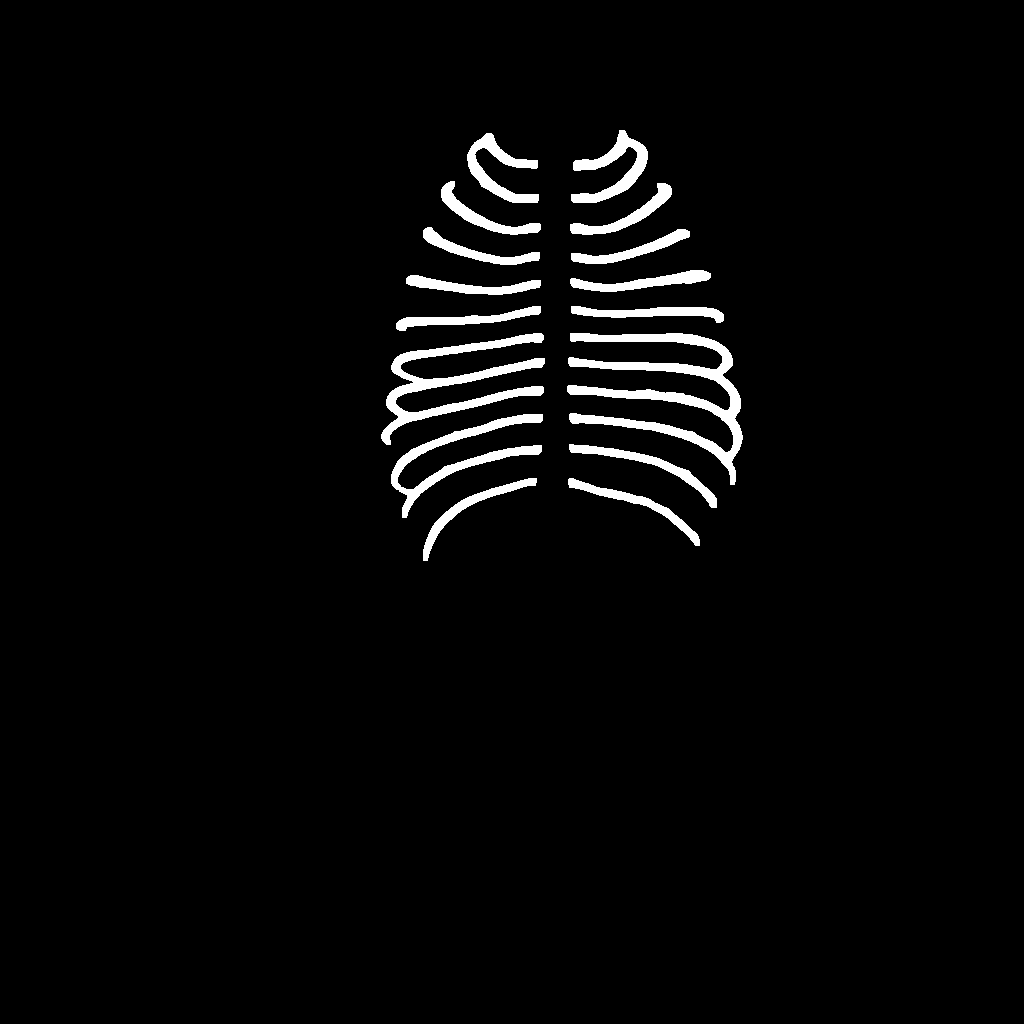}
        \caption{Ribs}
    \end{subfigure}
    \begin{subfigure}[b]{0.19\textwidth}
        \includegraphics[width=1.0\linewidth,height=1.2in]{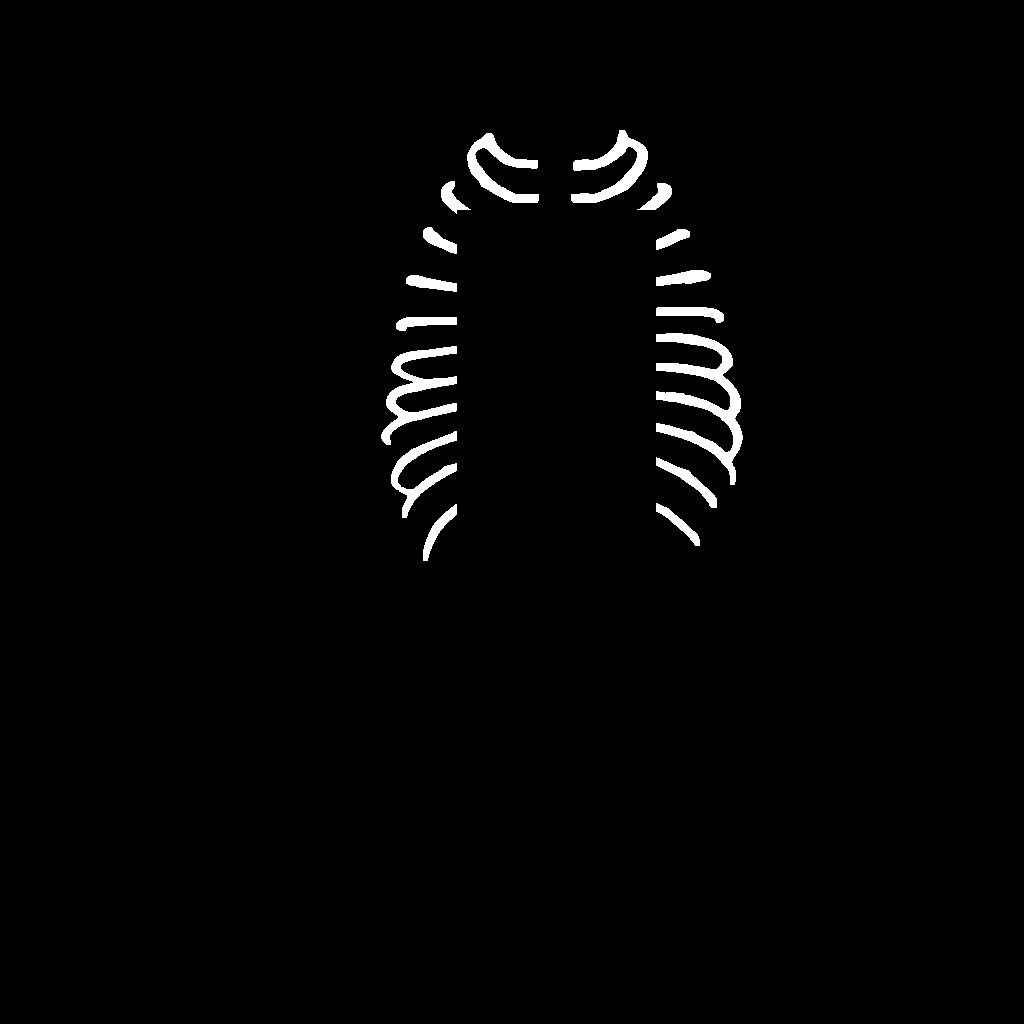}
        \caption{Obscured ribs}
    \end{subfigure}
    \begin{subfigure}[b]{0.19\textwidth}
        \includegraphics[width=1.0\linewidth,height=1.2in]{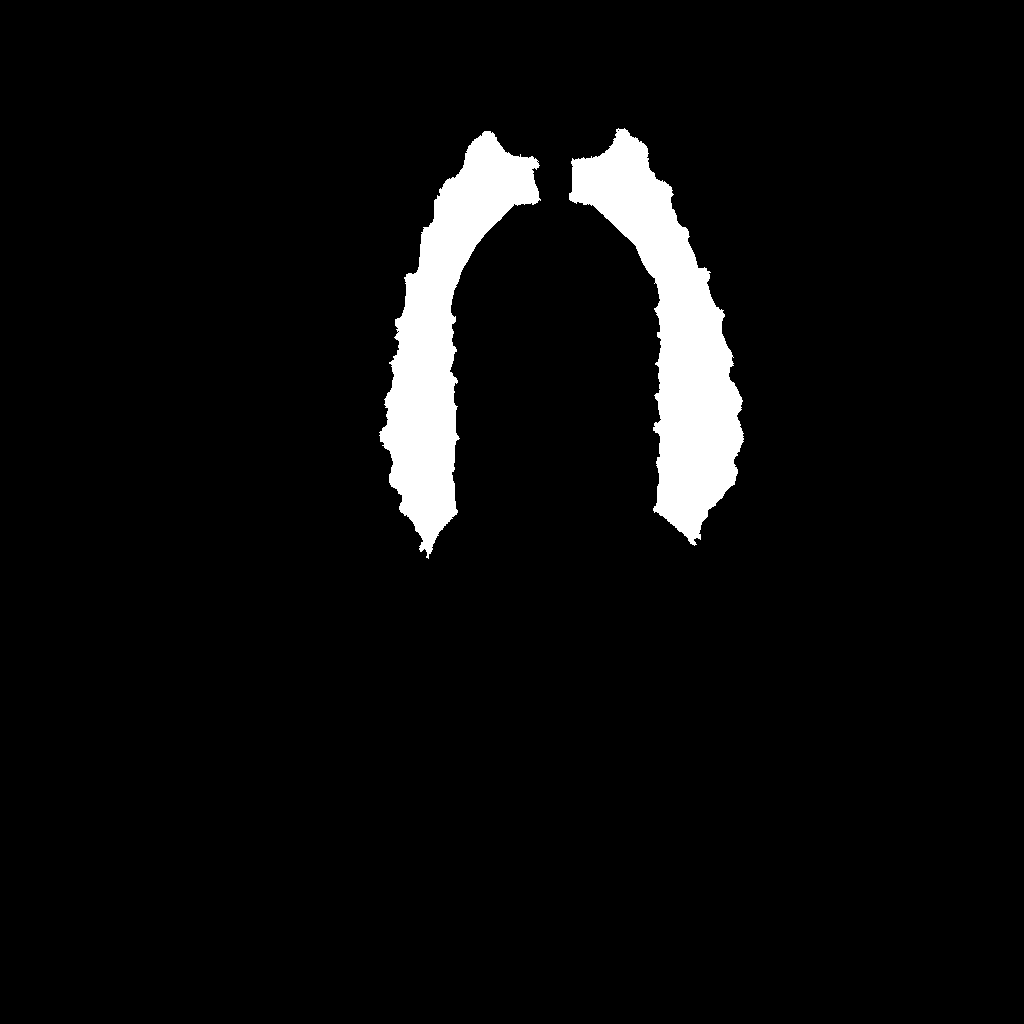}
        \caption{Two-snake result}
        \label{fig8-d}
    \end{subfigure}
    \begin{subfigure}[b]{0.19\textwidth}
        \includegraphics[width=1.0\linewidth,height=1.2in]{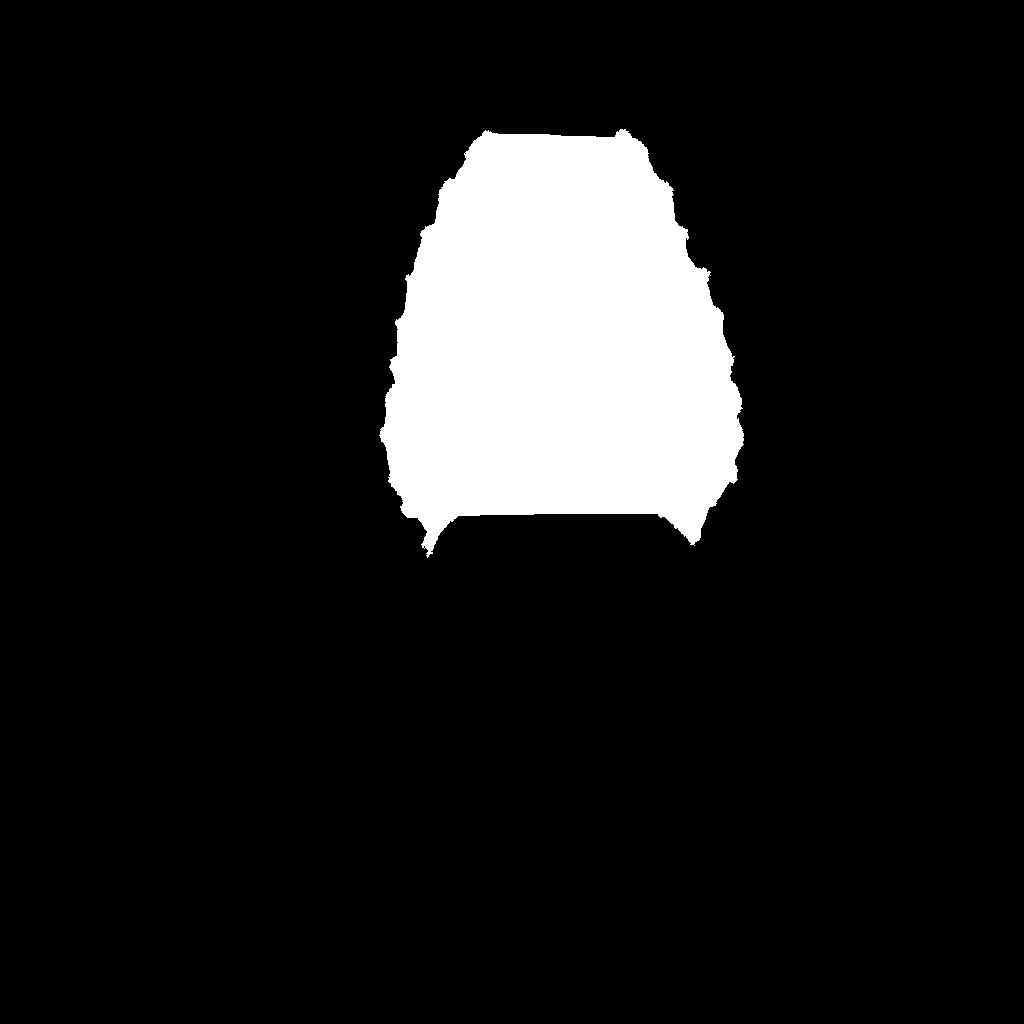}
        \caption{Our result}
        \label{fig8-e}
    \end{subfigure}

    \caption{Comparison between the results of our one-snake model and the two-snake method.}
    \label{fig8}
\end{figure}

\begin{table}[h]
\centering
\caption{Masking the left and right hemithoraces using two approaches: (1) the proposed method where the thorax was obtained by applying an active contour to the entire ribs, and the resulted region then was divided into the left and right hemithoraces, and (2) two-snakes method where the left and right ribs were treated by two active contours separately.}
\begin{tabular}{|l|l|l|}
\hline
\diagbox{Model}{Metric} & F1             & AUC           \\ \hline
(1) one-snake                   & \textbf{82.82} & \textbf{85.3} \\ \hline
(2) two-snake                   & 79.25          & 82.69         \\ \hline
\end{tabular}
\label{tb3}
\end{table}
Besides the computational efficiency of the proposed method, fitting active contour to all of the ribs can automatically fill any internal artifacts. This is investigated in Fig. \ref{fig8}, where the middle region of the ribs was manually obscured, and the hemithoraces were masked using the one-snake and two-snakes methods. Results of the two-snakes method showed that the segmented hemithoraces are significantly affected by the obstruction of the ribs, and this model could not fully recover the thorax shape, as can be seen in Fig. \ref{fig8-d}. However, our proposed model recovered the thorax shape, and missing rib regions did not affect the output, as shown in Fig. \ref{fig8-e}.
\section{Discussion and conclusion}

In this study, we introduced a novel thorax segmentation method for canine and feline VD and DV radiographs and utilized it to classify the symmetry of hemithoraces. The experimental results suggest that the proposed method for segmenting the thorax performs well under poor exposure settings and even when parts of the lungs are obscured. This ability came from the nature of fitting an active contour to a shape in which our snake is attracted to the image gradient and maintains the topology of the segmentation.

\begin{figure}[htb]
    \centering
    \begin{subfigure}[b]{\textwidth}
        \centering
        \includegraphics[width=0.3\linewidth, height=2in]{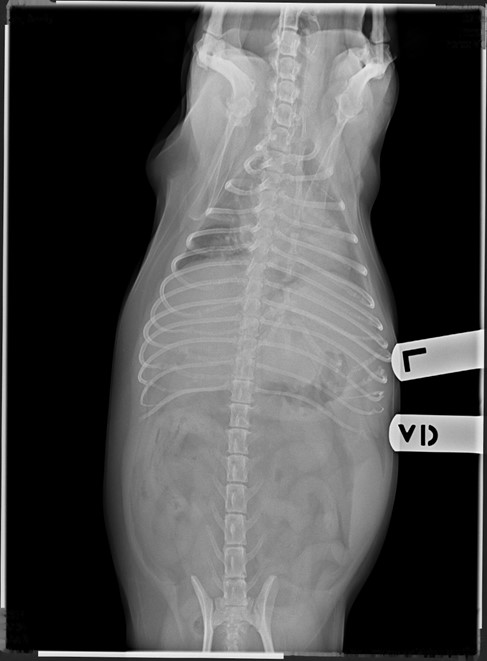}%
        \hspace{0.5in}
        \includegraphics[width=0.3\linewidth, height=2in]{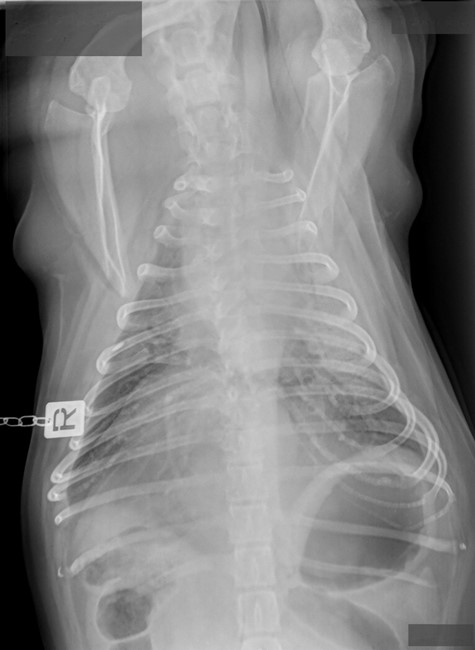}
        \caption{Example of radiographs with thoracic diseases causing lungs to appear bright.}
        \label{fig9-1}
    \end{subfigure}
    \vskip\baselineskip
    \begin{subfigure}[b]{\textwidth}
        \centering
        \includegraphics[width=0.3\linewidth, height=2in]{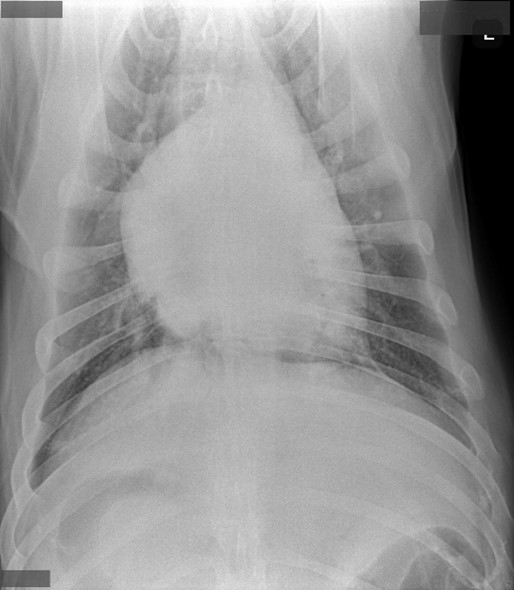}%
        \hspace{0.5in}
        \includegraphics[width=0.3\linewidth, height=2in]{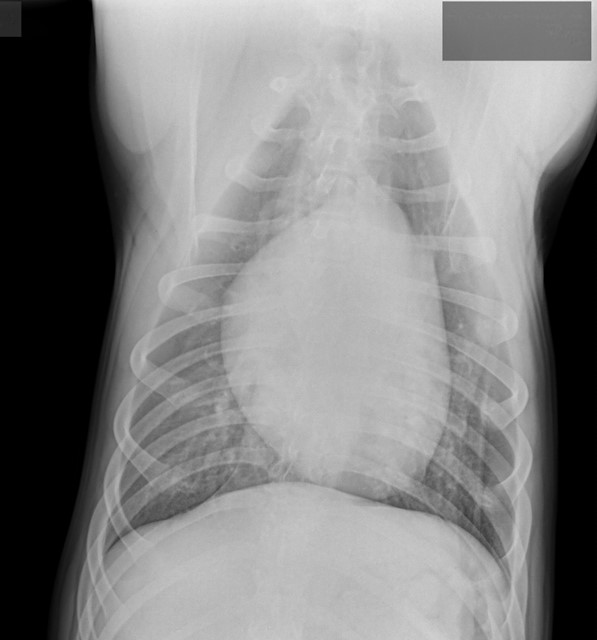}
        \caption{Examples of underexposed radiographs which cause lungs to appear bright.}
        \label{fig9-2}
    \end{subfigure}
    \caption{Examples of radiographs that the direct segmentation method would fail to segment the thorax. (a) The left lung lobes are affected by an increased opacity due to bronchopneumonia. (b) radiographs are underexposed, leading to a more white and grainy appearance of the pulmonary parenchyma.}
    \label{fig9}
\end{figure}

In thoracic radiography, the air-filled lungs are lucent regions (dark) relative to the soft tissue and bone within and surrounding the thoracic cavity. For symmetry classification, as studied by Santosh et al. \cite{11}, one possible solution is to label lungs based on their opacity (dark regions inside the thorax) in both VD and DV radiographs instead of segmenting the whole hemithoraces, as we proposed in this work. Then, use the masks to train an end-to-end segmentation model using these labels as supervision; we call this model a direct segmentation of lungs.  However, this method has two major drawbacks:

\begin{itemize}
\item Thoracic diseases can alter the expected opacity of the lungs.  For example, patients with pneumonia or heart failure will have an increased opacity of their lungs and patients with pleural effusion will have increased opacity in the pleural space.  Increased opacity in these regions may make it hard for the direct segmentation method to segment lungs due to the absence of the dark regions in the thorax that are typically indicative of lungs. Two examples of this scenario are represented in Fig. \ref{fig9-1}.
\item Lungs in underexposed radiographs may appear too bright or have increased noise. Two examples of underexposure are provided in Fig. \ref{fig9-2}.
\end{itemize}

An ensemble of MLP, SVM and GBC models allowed us to analyze and look at the features from different perspectives as a result of using different learning processes that improved the outcome. GBC, similar to a random forest, creates an ensemble of trees, each of which is weighed by a weight metric. The MLP uses a series of connected neurons, in which each layer transforms input data to a new space before making the final classification decision. The SVM tries to find a decision boundary between the points by first finding the distance between each data point using an RBF and then minimizing SVM loss. 
One limitation of our methods was that radiographs with an obscured spine were not tested in our model. However, since the spine's shape is relatively easy to estimate, even if a part of it is not visible, we can utilize the segmented part to recover an estimate of its full shape. Another drawback of the proposed method is that fitting an active contour might be time-consuming and is proportional to the resolution of the image. In fact, in active contour fitting, we have to consider all pixels and iteratively optimize the objective function. Thus a higher resolution radiograph with a higher number of pixels requires more steps to optimize the active contour. Evaluations on the test set showed that the average time to output an active contour on a radiograph of size 1024x1024 pixels is 3.0±0.5 seconds. 

In future works, we will utilize the proposed method for symmetry classification to develop an automatic quality classification for canine thoracic radiographs, where asymmetric left and right hemithoraces are an indication of poor collimation.  Therefore, the developed model can be incorporated into a larger model which analyzes and classifies the collimation of a given thoracic radiograph.

\subsection*{Disclosures}
The authors declare that there are no conflicts of interest related to this article.


\bibliography{report}   
\bibliographystyle{spiejour}   

\vspace{1ex}
\noindent Biographies of the authors are not available.

\end{spacing}
\end{document}